\documentclass[useAMS,usenatbib,bibyear]{aa}

\usepackage{txfonts}
\usepackage{epsfig,amsmath,mathrsfs}
\usepackage{natbib}
\bibpunct{(}{)}{;}{a}{}{,} % to follow the A&A style
\usepackage[dvips,usenames,dvipsnames]{color}
\usepackage{graphicx}
\usepackage[]{hyperref}
\usepackage{bm}
\usepackage{verbatim}
\usepackage{amssymb}

\newcommand{\moca}{{\sc MoCaLaTA}}
\newcommand{\uv}{UltraVISTA}
\newcommand{\lya}{Ly$\alpha$}

\definecolor{gray}{RGB}{180,180,180}

\renewcommand{\sec}[1]{Sec.~\ref{sec:#1}}
\newcommand{\fig}[1]{Fig.~\ref{fig:#1}}

\newcommand{\eq} [1]{Eq.~\ref{eq:#1}}

\newcommand{\Fig}[1]{Figure~\ref{fig:#1}}

\newcommand{\nHI}    {\ensuremath{n_{\textrm{{\scriptsize H}{\tiny \hspace{.1mm}I}}}}}

\newcommand{\NOVI}   {\ensuremath{N_{\textrm{{\scriptsize O}{\tiny \hspace{.1mm}VI}}}}}

\newcommand{\xHI}    {\ensuremath{x_{\textrm{{\scriptsize H}{\tiny \hspace{.1mm}I}}}}}

\newcommand{\X}      {\ensuremath{\mathbf{X}}}
\newcommand{\Y}      {\ensuremath{\mathbf{Y}}}

\newcommand{\xx}     {\ensuremath{X}}
\newcommand{\siglos} {\ensuremath{\sigma_{\!\scalebox{.5}{$-$}}}}
\newcommand{\sigloxs}{\ensuremath{\sigma_{\!\scalebox{.5}{$x-$}}}}
\newcommand{\sigloys}{\ensuremath{\sigma_{\!\scalebox{.5}{$y-$}}}}
\newcommand{\sighis} {\ensuremath{\sigma_{\!\scalebox{.5}{$+$}}}}
\newcommand{\sighixs}{\ensuremath{\sigma_{\!\scalebox{.5}{$x+$}}}}
\newcommand{\sighiys}{\ensuremath{\sigma_{\!\scalebox{.5}{$y+$}}}}
\newcommand{\siglo}  {\ensuremath{\sigma_{{\raisebox{-1pt}{\scriptsize \!$-$}}}}}
\newcommand{\sighi}  {\ensuremath{\sigma_{{\raisebox{-1pt}{\scriptsize \!$+$}}}}}
\def\ergs{\mbox{\,erg~s$^{-1}$}}
\def\Mpc{\mbox{\,Mpc}}
\def\hMpc{\mbox{\,$h^{-1}$\,Mpc}}
\def\hMpcs{\mbox{\,$h^{-2}$\,Mpc$^2$}}
\def\hMpcc{\mbox{\,$h^{-3}$\,Mpc$^3$}}
\def\hcMpc{\mbox{\,$h^{-1}$\,cMpc}}
\def\hcMpcs{\mbox{\,$h^{-2}$\,cMpc$^2$}}
\def\hcMpcc{\mbox{\,$h^{-3}$\,cMpc$^3$}}

\def\ergscm2{\mbox{\,erg~s$^{-1}$~cm$^{-2}$}}
\def\as2{\mbox{\,arcsec$^2$}}
\def\deg2{\mbox{\,deg$^2$}}
\def\cm2{\mbox{cm$^{2}$}}
\newcommand{\ave}[1]{\ensuremath{\langle #1 \rangle}}
\newcommand{\ten}[1]{\ensuremath{10^{#1}}}
\newcommand{\e}[1]  {\ensuremath{\times10^{#1}}}

\def\z{\mbox{$z = 8.8$}}                        %Redshift
\def\Msun{\mbox{$M_\odot$}}                     %Solar mass
\def\Mh{\mbox{$M_{\mathrm{h}}$}}                %Halo mass
\def\hmMsun{\mbox{$h^{-1}M_\odot$}}             %Msun/h
\def\rvir{\mbox{$r_{\mathrm{vir}}$}}            %Virial radius
\def\Mvir{\mbox{$M_{\mathrm{vir}}$}}            %Virial mass
\def\dMdt{\mbox{$\dot{M}_\mathrm{gas}$ }}       %Gas mass accretion rate
\def\OmegaM{\mbox{$\Omega_{\mathrm{M}}$}}       %Matter density parameter
\def\OmegaMz{\mbox{$\Omega_{\mathrm{M},z}$}}       %Matter density parameter
\def\OmegaL{\mbox{$\Omega_\Lambda$}}            %Dark energy density parameter
\def\OmegaLz{\mbox{$\Omega_{\Lambda,z}$}}            %Dark energy density parameter
\def\fesc{\mbox{$f_\mathrm{esc}$}}              %Escape fraction
\definecolor{DarkRed}{RGB}{195,0,0} % "RGB" -> [0,255]; "rgb" -> [0,1]
\hypersetup{colorlinks,%
            citecolor=OliveGreen,%
            filecolor=black,%
            linkcolor=DarkRed,%
            urlcolor=blue}%,%
\begin{document}

\title{Chasing Lyman $\bm{\alpha}$-emitting galaxies at $\bm{z = 8.8}$}
%\title[LAEs at \z\ with \uv]
%      {Chasing Lyman $\bm{\alpha}$ emitting galaxies at $\bm{z = 8.8}$}
    % {Chasing Lyman $\bm{\alpha}$ emitting galaxies at $\bm{z = 8.8}$\\
    %  --- is it worth the effort?}
\author{Peter Laursen
       \inst{1,2}
     % \fnmsep
       \and
       Jesper Sommer-Larsen
       \inst{2,3,4}
       \and
       Bo Milvang-Jensen
       \inst{2,5}
       \and
       Johan P. U. Fynbo
       \inst{2,5}
%\newauthor Alexei O. Razoumov$^6$ \\
       \and
       Alexei O. Razoumov
       \inst{6}
       }

       \institute{Institute of Theoretical Astrophysics, University of Oslo, P.O. Box 1029 Blindern, N-0315 Oslo, Norway.
                  \email{pela@astro.uio.no}
                  \and
                  Dark Cosmology Centre, Niels Bohr Institute, University of Copenhagen, Juliane Maries Vej~30, 2100, Copenhagen {\O}, Denmark.
                  \and
                  Marie Kruses Skole, Stavnsholtvej 29-31, DK-3520 Farum, Denmark.
                  \and
                  Excellence Cluster Universe, Technische Universit\"at M\"unchen, Boltzmannstra{\ss}e 2, 85748 Garching, Germany.
                  \and
                  Cosmic Dawn Center, Niels Bohr Institute, University of Copenhagen, Juliane Maries Vej 30, 2100 Copenhagen {\O}, Denmark.
                  \and
                  WestGrid, Vancouver, BC, Canada.
                  }

\date{\today}

\abstract{
With a total integration time of 168 hours and a narrowband filter tuned to
\lya\ emission from \z, the \uv\ survey has set out to
find some of the most distant galaxies, on the verge of the Epoch of
Reionization.
Previous calculations of the expected number of detected \lya-emitting galaxies
(LAEs) at this redshift based e.g.~on extrapolation of
lower-redshift luminosity functions did not explicitly take into account the
radiative transfer of \lya.
In this work we combine a theoretical model for the halo mass function,
i.e.~the expected number of haloes per volume, with numerical results from
high-resolution cosmological hydrosimulations post-processed with radiative
transfer of ionizing UV and \lya\ radiation, assessing the visibility of LAEs
residing in these haloes.
Uncertainties such as cosmic variance and the anisotropic escape of \lya\
are taken into account, and it is predicted that
once the survey has finished, the probabilities of detecting none, one, or more
than one are roughly 90\%, 10\%, and 1\%, respectively.
This is a significantly smaller success rate compared to earlier predictions,
due to the combined effect of
a highly neutral intergalactic medium (IGM) scattering \lya\ to
such large distances from the galaxy that they fall outside the observational
aperture, and to the actual depth of the survey being less than predicted.

Because the IGM affects narrowband (NB) and broadband (BB)
magnitudes differently, we argue for a relaxed colour selection criterion of
$\mathrm{NB} - \mathrm{BB} \simeq +0.85$ in the AB system.
Since the flux is dominated by the continuum,
however, even if a galaxy is detectable in the NB, its probability of being
selected as a narrowband excess object is $\lesssim35$\%.

Various additional properties of galaxies at this redshift are predicted,
e.g.~%
% UV/Lya LFs
the \lya\ and UV luminosity functions,
% SMHM
the stellar mass--halo mass relation,
% spec
the spectral shape,
% F_ap
the optimal aperture,
as well as
% fesc
the anisotropic escape of \lya\ through both the dusty, interstellar medium
and through the partly neutral IGM.
% For instance, we show that due to scattering in the partly neutral
% circumgalactic medium, the \lya\ radiation tends to escape at such large
% distances from the galactic centre that a standard circular aperture for sourc
% identification of diameter $2''$ will miss of the order of half the light.
% However, since larger apertures also mean more noise, using a larger aperture
% is not recommended; in fact the probability of detecting LAEs is highest
% for an aperture of roughly 1\farcs4, assuming a seeing representative for the
% \uv\ survey.

Finally, we describe and make public a fast numerical code for adding numbers
%with asymmetric uncertainties (``$x_{-\siglos}^{+\sighis}$'') which proves
with asymmetric uncertainties (``$x_{-\sigma_{-}}^{+\sigma_{+}}$'') which proves
to be significantly better than the standard, but wrong, way of adding upper
and lower uncertainties in quadrature separately.
}

\keywords{
radiative transfer --- galaxies: high-redshift --- cosmology: reionization
}

\maketitle

\section{Introduction}
\label{sec:intro}

%\gray{Remember to locate first instances of
%AMR,
%CGM,
%CI,
%cMpc,
%DLA,
%DM,
%EOR,
%GRB,
%HMF,
%IGM,
%IMF,
%ISM,
%LAE,
%LOS,
%PDF,
%QSO,
%RT,
%SN,
%SPH,
%UVB,
%and define these abbreviations (don't forget their plurals).}

Unveiling the evolution and the constituents of the early Universe is as
exciting and important as it is arduous.  While the formation of structure from
primordial fluctuations to massive haloes can be accounted for purely by
gravitational collapse of matter, the mechanisms that lead to the emission of
light --- the primary way of gaining information from the high-redshift
Universe --- are less clear. The first generation of stars seem to have emerged
some 200 Myr after the Big Bang \citep{Barkana2001,Visbal2012,Bowman2018}. In
a virtually neutral Universe, the escape of ionizing radiation was inhibited,
but eventually the Universe was reionized and galaxies became more easily
detectable.

Hard UV radiation is emitted from the most massive and hence hot stars of a
starburst, attempts to escape the surrounding neutral gas, and carves out bubbles
of ionized hydrogen in the interstellar medium (ISM). Because of the high
densities involved during the Epoch of Reionization (EoR), recombination
happens on very short timescales, eventually reprocessing 1/3 of the radiation
bluewards of 912 {\AA} into a single emission line --- the \lya\ line at 1216
{\AA}. Consequently, up to 10\% of the total, bolometric luminosity can be
emitted in \lya, as first predicted by \citet{Partridge1967}. At high
redshifts, where the metallicity may be expected to be substantially lower and
the initial mass function (IMF) may be more top-heavy than the
\citet{Salpeter1955}-one assumed by \citet{Partridge1967}, the fraction may be
even higher, up to 20--40\% \citep{Raiter2010}. This fortunate fact facilitates
observations of galaxies that would otherwise be impossible to detect.  
In addition to the \lya\ precipitated by stars, a notable amount is surmised to
be emitted from cooling radiation following the cold accretion of gas onto the
galaxies (see discussion in \sec{Lcool}).

When the \uv\ survey (discussed in detail in \sec{uv}) set out, the most
distant known galaxy was the \citet{Iye2006} \lya\ emitter (LAE) at
$z = 6.96$, which held its record for five years. Eventually,
a few LAEs at redshifts beyond 7 were reported
\citep[e.g.][]{Vanzella2011,Shibuya2012,Finkelstein2013}, and more recently,
galaxies at $z=8.68$ \citep{Zitrin2015}, $z=11.09$ \citep{Oesch2016},
and $z=9.1$ \citep{Hashimoto2018}
were detected, showing that the survey should at least have a chance of
detecting LAEs at \z.  
The galaxy detected by \citet{Oesch2016} was detected due to the \lya\
\emph{break} rather than the \emph{line}, while the one found by
\citet{Hashimoto2018} was detected in \ion{O}{iii} albeit with tentative
evidence for a \lya\ line; thus, an \uv-detection at \z\ could
be the highest-redshift LAE to date.

Detecting the most distant astronomical object has long been a goal in itself,
but pushing such observations to the limit is not just a question of increasing
the volume of the explored part of the Universe, or breaking a record. 
Although the difference in lookback time between, say, $z=8$ and 9 is merely a
hundred Myr, this is a significant fraction of the age of the Universe at this
time, which was only half a Gyr. Thus, these extreme observations offers
crucial insight into the time scales of the formation, evolution and general
properties of galaxies and the intergalactic medium (IGM).  Moreover, the number density and
clustering of LAEs at $z>6$ can put constraints on reionization
\citep{Furlanetto2006,McQuinn2007,Iliev2008,Jensen2013}. As yet, quantifying
the clustering of LAEs has been rather inadequate due to limited sample sizes,
but at these redshifts even the detection of a single galaxy provides important
constraints.  

\subsection{The \uv\ survey}
\label{sec:uv}

The comissioning of the 4.1 m survey telescope VISTA, located close to the VLT
on a neighbouring peak, and its wide-field infrared camera VIRCAM
\citep{Sutherland2015} was completed in 2009. 
VISTA is used mainly for large, multi-year, public surveys
\citep{Arnaboldi2017}.  One of these is
%\uv\footnote{PIs: J. Dunlop, O. Le F\`evre, M.  Franx, and J. P. U. Fynbo,
%\href{http://ultravista.org}{\tt{ultravista.org}}.}
\uv\ \citep{McCracken2012},
which utilizes the VISTA telescope to obtain deep imaging of $1.2\times1.5$
deg$^2$ of the COSMOS field \citep{Scoville2007} in the four near-IR filters
$Y$, $J$, $H$ and $K_\mathrm{s}$. Half of this field, the four so-called
ultra-deep stripes covering about $1\,\mathrm{deg}^2$, are also observed in a
narrowband filter NB118 centered at
1.19 $\mu$m \citep{Milvang-Jensen2013}, corresponding to \lya\ redshifted from
\z, or less than 600 Myr after the Big Bang.
This wavelength was chosen to be relatively free of airglow emission. 
Unfortunately the filter turned out to be subject to a small shift in
wavelength of approximately 3.5 nm when mounted in VIRCAM, which, while still
effectively probing the same redshift, is large enough that airglow lines are
partly entered, resulting in a somewhat lower sensitivity
\citep{Milvang-Jensen2013}.  

The narrowband survey, which constitutes $\sim10\%$ of the full \uv\ survey,
set out in December 2009 to construct an ultradeep image from 168 hours of
integration.
An original 5$\sigma$ detection limit of $F=3.7\e{-18}\ergscm2$ (corresponding
to an AB magnitude of 26.0) was expected \citep{Nilsson2007}; however, due to
the above-mentioned shift, as well as a
too optimistic sky background assumed by the ESO Exposure Time Calculator used
at the time, it turned out to be
somewhat shallower: Based on the GTO data \citep{Milvang-Jensen2013}, the
detection limit for the final \uv\ is predicted to be
$F=1.5$--$1.7\e{-17}\ergscm2$ \citep[see e.g.][]{Matthee2014}.
The depth and the large area still results in a volume much larger than
previous surveys.

%\gray{The detection limits of \uv\ are so low that objects more than twice fainter
%than the \citeauthor{Iye2006}-LAE --- were it located at \z\ --- will be
%detected, provided they exist.}
%

The observational part of the narrowband survey was completed in January this
year.  Since the data reduction is extremely time-consuming, the latest data
release (``DR3'') only contains 60\% of the final narrowband exposure; so far
there is no sign of the detection of \z\ LAEs.
Can we hope to detect any at all? This question boils down to
estimating the luminosity function (LF) of LAEs at a hitherto unexplored epoch
of the history of the Universe. In essence, two complementary approaches exist
to this problem: Either one can look at LFs of galaxies probed at lower
redshift, and then extrapolate to \z, assuming a model for the evolution of
both galaxies and the IGM. Alternatively, one may resort to numerical
simulations, relying on the faith in our understanding of the physics governing
the formation of structure.  

\subsection{Previous predictions and observations at \z}
\label{sec:prevobs}

Prior to the launch of \uv, \citet{Nilsson2007} made predictions for the
expected number of detected LAEs at \z. At this time, realistic \lya\ radiative
transfer (RT) was at its infancy and unavailable to most astronomers. Instead
these authors followed three different approaches: Model 1 used the
semi-analytical galaxy formation model GALFORM \citep{Cole2000} with a common
\lya\ escape
fraction of $\fesc = 0.02$ or 0.2 for galaxies of all masses, to match LF
observations at lower redshifts \citep{Delliou2006}.
Model 2 uses the phenomenological model of \citet{Thommes2005}, normalized to
match the observed mass function of galaxy spheroids at $z = 0$. Model 3 uses a
Schechter LF with parameters extrapolated linearly from fits at $3.1 \le z \le
6.5$. The three models predict between 3 and 20 detections.

With similar extrapolations of lower-redshift LFs, \citet{Faisst2014} predicted
a less optimistic $0.6\pm0.3$, while \citet[][see below]{Matthee2014} predicted,
with their most optimistic LF, a number of $\lesssim23$. However, using the
\citet{Nilsson2007} LF extrapolation but with the
realistically estimated depth yields only 0.19.

% Common to the
% three methods is that they in some way are calibrated with results from lower
% redshifts where we do have detections, and then assume some evolution from higher
% redshift.
% As is common to all serious predictions, these methods assume an evolution of
% observables, calibrating unknown parameters to make them match observable
% objects. 
% However, as we are dealing with epochs for which not only LAEs have not yet been
% discovered, but for which \emph{any} observations are extremely sparse, we are
% really 

Observations of LAEs at \z\ have been attempted before.
\citet{Willis2005} used the VLT/ISAAC narrowband to sample a volume of 340
\hcMpcc\ to a limiting flux of $3.28\e{-18}\ergscm2$, while
\citet{Cuby2007}, using the same instrument, went for a larger area of 31
arcmin$^2$ albeit to a shallower threshold of $\sim1.3\e{-17}\ergscm2$.
Taking advantage of the lensing effect of three clusters, \citet{Willis2008}
reached $3.7\e{-18}$ in an area of 12 arcmin$^2$.
\citet{Sobral2009} surveyed $1.4\,\mathrm{deg}^2$ down to $\sim7.6\e{-17}
\ergscm2$ with UKIRT/WFCam in the HizELS survey at $z\simeq9$, and
\citet{Matthee2014} searched the hitherto largest area of $10\,\mathrm{deg}^2$
with CFHT/WIRCam and VLT/SINFONI followup down to $7\e{-17}\ergscm2$.
However, all surveys returned null-results.

%end prevobs

%end uv

\subsection{Overview of our approach}
\label{sec:overview}

Predicting numerically the expected the number of observed galaxies is a
challenging task, but can essentially be divided into two subprojects:
Calculating the number of galaxies in the cosmic volume surveyed by \uv, and
calculating the observability of those galaxies. In principle, everything could
be computed in a single numerical simulation calculating the galaxy formation
and emission of light from the galaxies. In practise, however, the simultaneous
need for a large volume and high resolution obliges us to split up the project
in several parts. In the following, the various constituents of the project
are outlined.  

Galaxies reside in dark matter haloes, the number density of which is given by
the \emph{halo mass function} (HMF) which can be estimated analytically or
determined via large cosmological simulations, or using a combination hereof.
Whether or not a given halo hosts a galaxy is given by a halo occupation
distribution, which is defined once the concept of a galaxy is defined. The
easiest approach is to simply define a galaxy as \emph{any} conglomerate of
baryons, or even any dark matter (DM) halo, whether or not these particles have
made themselves visible by producing any kind of luminous matter. Subsequently
we can then ask the question ``Is a given galaxy observable?''
% 
% However, not
% every halo hosts a galaxy, and some haloes host more than one. The relevant
% probability is given by the \emph{halo occupation distribution} (HOD).

Simulating galaxy formation realistically requires not only inclusion of
hydrodynamics, but also higher resolution than what is offered by large-scale
simulations. Moreover, for the RT \emph{much} higher resolution is needed,
making it impossible to simulate a statistically robust sample
of galaxies with sufficient resolution. These difficulties are dealt with
by carrying out a fully cosmological, hydrodynamical, but medium-sized,
simulation which is large
enough that the interaction between individual galaxies is accounted for.
%yet small enough that the ISM of the galaxies is sufficiently resolved for the RT.
Subsequently, a number of galaxies are extracted from the simulation and
resimulated at higher resolution. Several different models for star formation
and feedback are investigated.

Finally, two RT schemes are employed: While the cosmological simulation uses an
on-the-fly approximation of the ionizing UV background (UVB), a full UV RT is
performed on the extracted galaxies and their environments, after which the
\lya\ RT is conducted, monitoring the spatial and spectral distribution of the
photons that reach the observer, i.e.~those which are not absorbed by dust, or
scattered out of the line of sight (LOS) by the IGM.

The two subprojects --- calculating the number density and the observability
--- are described in Secs.~\ref{sec:n_gal} and \ref{sec:obs}, respectively.
The results are presented in \sec{res}, discussed in \sec{disc}, and
summarised in \sec{sum}.

%end overview

          %%%%%%%%%%%%%%%%%%%%%%%%%%%%%%%%%%%%%%%%%%%%%%%%%%%%%
          %                                                   %
          %   Somewhere:  Hayes 10, Dijkstra & Jeeson 13      %
          %   and Forero define f_0cc as f_obs                %
          %                                                   %
          %%%%%%%%%%%%%%%%%%%%%%%%%%%%%%%%%%%%%%%%%%%%%%%%%%%%%

%end intro

\section{Galaxy number density}
\label{sec:n_gal}

\subsection{Halo mass function}
\label{sec:HMF}

The problem of calculating the distribution of the masses \Mh\ of collapsed
haloes was addressed first by \citet[PS;][]{Press1974}, who considered the
spherical collapse of gravitationally interacting matter from an initial,
smoothed density field.
The resulting HMF expressed as number density per logarithmic mass bin can be
written as
\begin{equation}
\label{eq:HMF}
\frac{dN}{d\ln\Mh} = \frac{\rho_{\mathrm{M,0}}}{\Mh}
                    \left| \frac{d\ln\sigma}{d\Mh} \right|
                    f(\sigma),
\end{equation}
where
$\rho_{\mathrm{M,0}}$ is the present-day average mass density of the
Universe,
$\sigma(\Mh,z)$ is the rms fluctuations of the density field smoothed
with a top-hat filter $W(R)$ of radius
$R = (3\Mh / 4\pi\rho_{\mathrm{M}}(z))^{1/3}$, and
$f(\sigma)$ is discussed below.

For a linearly extrapolated density field characterised by a power
spectrum $P(k)$, the rms fluctuation is given by
\begin{equation}
\label{eq:sigmaM}
\sigma^2(M) = \frac{\delta^2(z)}{2\pi^2}
              \int_0^\infty
              dk \, k^2 P(k) \widehat{W}(k,M),
\end{equation}
where $\widehat{W}(k,M) = 3 (\sin kR - kR\cos kR) / (kR)^3$ is the Fourier
transform of the real-space filter\footnote{Note that other functional forms of
the filter are possible, e.g. a Gaussian filter for which $\widehat{W}(k,M) =
e^{-(kR)^2/2}$, where the relation between mass and radius is $\Mh =
(2\pi)^{3/2} R^3 \rho_\mathrm{M}$.}
$W(R)$, and $\delta(z) = D(z)/D(0)$ is
the linear growth rate normalized to unity today. The growth rate function, in
turn, involves an integral over the expansion history, but is very well
approximated \citep{Lahav1991,Carroll1992} by
\begin{equation}
\label{eq:D}
D(z) = \frac{5}{2}
       \frac{1}{1+z}
       \frac{\OmegaMz}
       {\OmegaMz^{4/7} - \OmegaLz + (1+\OmegaMz/2) (1+\OmegaLz/70)}
\end{equation}

The last term in \eq{HMF}, $f(\sigma)$, is the \emph{multiplicity function}
giving the mass fraction associated with halos in a unit range of $\ln\nu$,
with $\nu \equiv \delta_\mathrm{cr}/\sigma$ and $\delta_\mathrm{cr} \simeq
1.686$ the critical density contrast needed for gravitational collapse.
Only in the PS formalism is an
analytical form of $f(\sigma)$ obtainable; in general it must be empirically
derived by fitting to halo abundances in cosmological simulations. 
The PS HMF gave good fits to simulations at that time,
but as computing resources improved, it was found numerically that it
over(under)predicts the collapsed fraction at the low-(high-)mass end
\citep[e.g.][]{Governato1999}.

By introducing two extra parameters,
\citet[ST;][]{Sheth1999} gave an improved formalism (further detailed in
\citet{Sheth2001} and \citet{Sheth2002}), allowing for ellipsoidal structures
to form.
In the ST formalism the multiplicity function is given by
\begin{equation}
\label{eq:fsigST}
f(\sigma) = A \sqrt{\frac{2\tilde{\nu}}{\pi}}
              \left[ 1 + \tilde{\nu}^{-p} \right]
              e^{-\tilde{\nu}/2},
\end{equation}
where $A\simeq0.322$ is a normalization constant,
$\tilde{\nu} \equiv a\nu^2$ with
$a = 0.707$ describing a high-mass cutoff in an $N$-body simulation, and
$p = 0.3$ is given by the shape of the HMF at low masses in the simulation.
The PS HMF is recovered using $\{A,a,p\} = \{\frac{1}{2},1,0\}$).

While the ST HMF is much more accurate,
especially at low redshifts, it still tends to overpredict the abundance of
haloes at higher redshifts, when comparing to large, cosmological simulations.
On the basis of the so-called Bolshoi simulation --- an 8.5 billion particle
simulation of comoving volume $(250\, h^{-1}\mathrm{Mpc})^3$ ---
\citet{Klypin2011} provide a simple $z$-dependent correction factor which
brings the analytical predictions much closer to the results of the simulation:
\begin{equation}
\label{eq:Fdelta}
F(z) = \frac{[5.501\delta(z)]^4}{1 + [5.500\delta(z)]^4},
\end{equation}
The Bolshoi simulation uses a (near-)\emph{WMAP} seven-year data cosmology
\citep{Jarosik2010}. For our analysis we adopt the ST HMF with the correction
factor in \eq{Fdelta}; in \sec{HMFdisc} we discuss the implications of using
other HMFs and the more recent Planck cosmology
\citep{PlanckCollaboration2016}.

%end HMF

\subsection{\uv\ area and depth}
\label{sec:vol}

To get the total number of galaxies, the number density must be multiplied by
the volume surveyed by \uv. However, %while the volume itself is well-defined,
not all parts of the area are exposed equally. VIRCAM
comprises 16 detectors (each with its own NB118 filter)
which cannot be placed adjacently but instead are
separated by a large fraction of a detector size. To acquire a contiguous
image, the camera is shifted in the declination direction between exposures
(or ``paw prints''), resulting in most of the area being exposed twice, while
the top and bottom $5.5'$ are exposed once, in units of a single paw print
exposure time. With a final, total exposure time of 168 hours, split equally in
the three paw prints, most of the area will reach an exposure time of 112
hours.  
Moreover, detector \#16 and, to a lesser extend, \#4 have sizable unreliable
regions that are excluded from the analysis.
The result is that various parts of the area is sensitive to different
depths.  For details on exposures, see
\href{http://www.vista.ac.uk} {{\tt www.vista.ac.uk}}.  The total
(non-dead) area exposed at least once is 1.07 \deg2, which at \z\
corresponds to 150\hMpcs, or $14.5\e{3}\hcMpcs$.

The depth of the volume is also not unequivocal; since the
transmission curves of the filters are not tophats, different
distances are probed to different depths. Moreover, the 16 filters are slightly
different, and have slightly different central wavelengths. In general,
they all have roughly ``full'' transmission (i.e.~around 60\%) in a
window of $\sim 70$ {\AA}. Outside of this region the transmission tapers off
in such a way that the average FWHM is $123\pm3$ {\AA}.
If the emission lines were delta functions, at \z\ this would translate into
galaxies within a slab 19 \hcMpc\ thick being detected. 
Since in fact resonant scattering broadens the emission lines by several
{\AA}ngstr\"om (in the rest frame), galaxies on the edges of this slab are
less conspicuous than more centrally located ones. For an extensive description
of the filters and transmission curves, see \citet{Milvang-Jensen2013}.  

The consequence is that the exact volume becomes a function of the flux limit
of the observation; whereas a faint galaxy may only be detected if
it lies at the centre of a filter, a bright galaxy may be detected even if it
lies in the wing.  Thus, any survey probes bright galaxies in a larger volume
than fainter. An approximate volume can be computed from the average
filter FWHM and the area exposed at least once; this estimate yields 290
\hMpcc, or $2.7\e{5}$ \hcMpcc.

Likewise, line broadening from RT effects may make galaxies both more and less
visible: A \lya\ line that \emph{would} be detectable if it were a delta-like
line can be smeared out over tens of {\AA}ngstr\"oms (in the observer's frame),
lowering the flux density below the sensitivity threshold. On the other hand,
a (bright) line lying so far out in the wing of the filter that it would be
undetectable as a delta-line, could be broadened into the sensitive part of the
filter.

To address these complications, in the following calculations we 
integrate the simulated spectra over the exact filter shapes
%(i.e. shift the spectra slightly in redshift space over the full width of the
%filters, in steps of $\Delta z = 0.01$),
for each of the 16 filters, and split the detector up in 52 regions with
individual limiting magnitudes.

%end vol

%end n_gal

\section{Galaxy observability}
\label{sec:obs}

Whether a galaxy is observable depends on many factors, from its history of
formation, to its ability to create luminous objects, to the emitted light
making its way out through the galaxy and down to us. To simulate this
numerically requires detailed modelling; in particular high resolution and a
realistic HI field is needed for the \lya\ RT. In the following, the
simulations of the galaxies and the RT is described.

\subsection{Cosmological hydrosimulations}
\label{sec:hydro}

\subsubsection{The code}
\label{sec:simcode}

The TreeSPH type code used to simulate the formation and evolution of galaxies
to \z\ is in most aspects identical to the code described by
\citet{Sommer-Larsen2017}. In particular, the superwind prescription for
starburst-driven energy feedback is implemented, and a \citet{Chabrier2003} IMF
is employed. This enables the continuing formation of starburst-driven
(metal-enriched) outflows (discussed further in \sec{outflows}),
in line with indications
of recent observations of \ion{O}{vi} and \ion{O}{vii} absorption in the
Galactic CGM. Such outflows are also of great significance to the \lya\ RT
(see \sec{outflows}), especially during the
EoR \citep{Dijkstra2010,Dijkstra2011}.  The code used for the present
work differs from the one described in \citet{Sommer-Larsen2017} in a few
aspects: \emph{i}) a ``stretched'' version of the \citet{Haardt1996} UVB is
used, rather than the \citet{Haardt2012} UVB --- the stretched UVB turns on at
$z=12$, whereas the original \citet{Haardt1996} UVB turns on at $z=6$
(see \citealt{Laursen2011} for more detail), \emph{ii}) early
stellar feedback (ESF) was not implemented in the simulations described in this
work, and \emph{iii}) the effects of self-shielding of the UVB by gas is not
described using the mean field approximation of \citet{Rahmati2013}. Instead it
is assumed that the UVB is fully shielded off in regions where the mean free
path of Lyman limit photons is less than 0.1 kpc.

%end simcode

\subsubsection{The simulations}
\label{sec:sims}

% As will be discussed later, one
One
expects the brightest and most  \lya-luminous
galaxies observable at \z\ to be located in higher-density, proto-cluster
regions of the Universe. Hence, emphasis is placed on simulating
such regions at high resolution. To this end, the well-known ``zoom-in''
technique \citep[e.g.][and references therein]{Navarro1994,Onorbe2013} is
applied in two steps, as will be detailed in the following.

First, a low resolution, DM-only simulation is run to $z=0$, using $128^3$
particles in a $(150\hcMpc)^3$ box with periodic boundary conditions, and
starting at $z_i=39$. The cosmological parameters of the simulation (and the
subsequent simulations described below) are
$\{\OmegaM,\Omega_\Lambda,h,n,\sigma_8\} = \{0.3,0.7,0.7,0.95,0.9\}$.
This is somewhat different from the cosmology of the HMF, but we will only use
these simulations to assess the link between DM and baryons, and will moreover,
unless otherwise stated, factor out the $h$-dependence from cosmological
quantities.

Two of the most massive DM haloes, at $z=0$, are selected for re-simulation
using the zoom-in technique; they correspond in several aspects to the Virgo
and (sub)Coma clusters, with virial masses at $z=0$ of about $3\e{14}$ and
$1.2\e{15}$ \Msun\ and ``temperatures'' of $kT \sim 3$ and 6 keV, respectively
(see \citealt{Sommer-Larsen2005,Romeo2006} for more detail). For each cluster,
all DM particles within the virial radius at $z=0$ have been traced back to the
initial conditions at $z_i$. The particles are contained within ``virial
volumes'' having a linear extent of $20\text{--}30\hcMpc$, and within these
volumes the numerical resolution is increased by 512 times in mass, and eight
times in linear resolution. In addition, each higher resolution DM particle is
split into an SPH particle and a DM particle --- a baryonic fraction
$f_\mathrm{b} = m_\mathrm{SPH} / (m_\mathrm{DM}+m_\mathrm{SPH}) = 0.15$ is
assumed, in quite good agreement with recent values reported by the WMAP and
Planck collaborations.
 
The mass resolution reached inside the resampled Lagrangian sub-volumes is
$m_\mathrm{DM} = 2.2\e{8} \hmMsun$ and $m_\mathrm{SPH} = m_\star = 3.9\e{7}
\hmMsun$. Using the code described in \sec{simcode},
the evolution of the two virial volumes and the lower-resolution regions
around these is then simulated from $z_i=39$ to \z. The gravity softening
lengths $\epsilon$ are kept fixed in comoving coordinates until a redshift of
15, and at later times they are constant in physical coordinates. In the
higher-resolution regions, at $z<15$, $\epsilon_\mathrm{SPH} = \epsilon_\star =
0.75 h^{-1}\mathrm{kpc}$ and $\epsilon_\mathrm{DM} = 1.3 h^{-1}\mathrm{kpc}$.

At \z, 20 galaxies of stellar mass $M_\star \gtrsim 3\e{8} \Msun$,
corresponding to an absolute UV magnitude at 1600 \AA\ of $M_\mathrm{UV}
\lesssim -20$ (see \fig{MUV}), are selected from the two simulations. The
galaxies are identified in the higher-resolution regions of the two
proto-cluster simulations using the algorithm described, e.g., in
\citet{Sommer-Larsen2005}.  For each of the 20 galaxies, the DM particles
inside of the virial radius (at \z) are traced back to the initial conditions
at $z_i$ of the two resolution-increased proto-cluster simulations described
above. It is found that, in each case, this virial volume is contained within a
sphere of radius $3\hcMpc$.  For each galaxy, the numerical resolution in this
sphere (which is located within the higher-resolution region of the
corresponding proto-cluster initial conditions) is increased (additionally) by
64 times in mass, and four times in linear resolution. Consequently, the mass
resolution reached inside these resampled spheres is $m_\mathrm{DM} = 3.4\e{6}
\hmMsun$ and $m_\mathrm{SPH} = m_\star = 6.1\e{5} \hmMsun$.  

The 20 sets of nested, twice resolution-increased initial conditions form
the base of the ``production'' simulations presented in this work. For each of
the 20 sets, the code described in \sec{simcode} is used to simulate the
evolution of the formation region of a fairly massive galaxy, from $z_i = 39$
to \z. For computational reasons, the 20 high-resolution (HR) regions are not
evolved at the same time as one simulation, but are rather simulated
one-by-one, resulting in a total of 20 production simulations. At $z<15$,
$\epsilon_\mathrm{SPH} = \epsilon_\star = 188h^{-1}$pc
and $\epsilon_\mathrm{DM} = 335
h^{-1}$pc in the HR regions. At \z, the HR regions reach out to distances of
about 150 kpc (physical) from the central galaxies. The virial radii of the
central galaxies range from 12--25 kpc, so the HR regions around the galaxies
stretch 6--12 times as far as the virial radius. Consequently, other (typically
smaller) galaxies are also present in these 20 simulations. Using
the galaxy detection algorithm referred to above, a total of about 500 galaxies
are identified --- most of the analysis presented in this work builds on this
sample of simulated galaxies.
%the galaxy detection algorithm referred to above, a total of 268 ``primary'' galaxies are identified --- primary meaning that only the largest galaxy in a given DM halo is used for further analysis.

In addition to the 20 simulations discussed above, for convergence study
purposes, seven of the 20 galaxies were also simulated at eight times higher
mass resolution and twice better linear resolution. The initial conditions for
these seven simulations were constructed in the same way as described above,
except that (for computational purposes) each galaxy was represented by a
radius $1\hcMpc$ sphere in the initial conditions (rather than the radius
$3\hcMpc$ spheres described above), in which the mass resolution is increased
by 512 times (rather than the 64$\times$ above), and the linear resolution was
increased by a factor of eight (rather than the 4$\times$ above). The seven
simulations were then run with the code described above, from $z_i=39$ to \z. 
The mass resolution reached inside these resampled spheres is, consequently,
$m_\mathrm{DM} = 4.3\e{5} \hmMsun$ and $m_\mathrm{SPH} = m_\star = 7.6\e{4}
\hmMsun$.  Moreover, at $z<15$, $\epsilon_\mathrm{SPH} = \epsilon_\star = 94$
and $\epsilon_\mathrm{DM} = 167 h^{-1}$pc in the (ultra-)HR regions. At \z, the
HR regions only reach out to a couple of virial radii from the central
galaxies, but the simulations can nevertheless be used for some aspects of
convergence studies, detailed in App.~\ref{app:conv}.

%end sims

%end hydro

\subsection{Ionizing UV radiative transfer}
\label{sec:UVRT}

As described above,
the hydrosimulations employ an approximate scheme for the RT of ionizing UV
radiation. Beginning at $z_{\rm re} = 12$, a UVB field similar in shape to
\citet{Haardt1996} is assumed to build up, exposing all regions characterised
by a Lyman-limit photon mean free path greater than 0.1 kpc fully to the field,
while the remaining regions are assumed to be self-shielded. This results in
roughly half of the hydrogen in the simulation being ionized at $z\sim 10$--11,
consistent with what is inferred from WMAP polarization maps
\citep[][$z_{\rm re} = 10.6\pm1.0$]{Bennett2013}, but $\sim100$ Myr earlier
than the more recent result by 
\citet[][$z_{\rm re} = 8.8_{-1.4}^{+1.7}$]{PlanckCollaboration2016}.

To model the local UV coming from stars in the galaxy, we post-process the
snapshots at \z\ with a detailed UV RT scheme. The physical properties of the
SPH particles are first interpolated onto an adaptively refined grid of base
resolution $128^3$, with up to 15 additional levels of refinement, such that no
cell contains more than one particle. The smallest cells thus have a linear
extent $20\hMpc/(1+8.8)/128/2^{15} = 0.7$ pc. The same adaptive mesh refinement
(AMR) is used in the
subsequent \lya\ RT.  At each stellar source in the simulation we start twelve
rays covering all $4\pi$ directions isotropically which are then split
recursively into four rays each to maintain adequate resolution (at least
several ray segments per every cell in the volume) as we move further away from
the source or enter refined cells. Along these rays, the equlibrium
photoionization state of each cell is calculated iteratively taking into
account both transfer of stellar UV photons and the above-mentioned UVB.  The
stellar UV RT is computed separately in the frequency bands $[13.6,24.6]$ eV,
$[24.6,54.4]$ eV, and $[54.4,\infty]$ eV. For details, see
\citet{Razoumov2006,Razoumov2007}.  

%end UVRT

\subsection{\lya\ radiative transfer}
\label{sec:LyaRT}

\subsubsection{Galactic radiative transfer}
\label{sec:galRT}

To conduct the \lya\ RT we use the code \moca\
\citep{Laursen2009a,Laursen2009b}. \moca\ is a ray-tracing Monte Carlo code
that operates on an AMR grid, so we use the same grid that was constructed for
the UV RT. In the following, the characterstics of the code are briefly
outlined: 

Each cell contains the physical variables important for the RT, which are \lya\
and continuum emissivity $L_{\mathrm{Ly}\alpha}$ and $L_{\mathrm{FUV}}$, gas
temperature $T$, velocity field $\mathbf{v}_{\mathrm{bulk}}$, as well as
densities \nHI\ and $n_{\mathrm{d}}$ of neutral hydrogen and dust.  \lya\
emissivity is calculated as the sum of \lya\ photons produced from
recombinations following ionization (mainly from hot stars, but also from the
UVB field) and cooling radiation, discussed in more detail in the next
section.  

Individual photons are emitted from a given cell with a probability
proportional to the emissivity in that cell, in a random direction, and is
followed as it scatters stochastically on the neutral hydrogen out through the
ISM, until it either escapes the galaxy or is absorbed by dust. The position
and wavelength of the photons are recorded in an IFU-like array along the
Cartesian axes, i.e.~the galaxy is observed simultaneously from six different
directions. This provides a measure of the anisotropic escape of \lya\ from the
galaxies.

%end galRT

\subsubsection{Cooling radiation from cold accretion}
\label{sec:Lcool}

As mentioned in the previous section, accretion of gas onto galaxies is
expected to give rise to some emission of \lya\ photons in addition to the
photons resulting from star formation.
Exactly how much is a matter of debate, since numerical
predictions \citep[e.g.][]{Goerdt2010,Faucher-Giguere2010} require
accurate knowledge of the thermal state of the gas, which is difficult, and
since observations are lacking or indirect (see discussions in
\citealt{Dijkstra2014} and \citealt{Prescott2015}). As a (semi-)analytical
estimate showing that at least it should not be readily neglected, we can take
cooling radiation $L_{\mathrm{Ly}\alpha\mathrm{,cool}}$ to be proportional to
the gas mass accretion rate \dMdt\ and the potential difference $|\Delta\Phi|$
from the gas begins to accrete till it settles at the core.

Free-falling gas will deposit its released energy in bulk kinetic form.
In contrast, gas falling at constant velocity will heat and subsequently cool.
Taking the ambient temperature of the halo gas to be equal to the virial
temperature and the velocity of gas in the cold streams to be equal to the
sound speed implies that the gas falls at a constant velocity roughly equal to
the virial velocity. In this case, all the energy released
will go into heat, and one proportionality factor will be $f_\gamma \simeq 1$.
This approximation tends to be accurate especially at high redshifts
\citep{Goerdt2015}.
A more conservative choice might be $f_\gamma=0.3$ \citep{Dijkstra2009}.
The amount of radiation emitted as \lya\ depends on the
temperature of the gas, but as discussed in \citet{Fardal2001}, the vast
majority of the cooling radiation comes from gas with $T\sim10^4\,\mathrm{K}$,
implying that a fraction $f_\alpha \sim 0.5$ is emitted as \lya.  

The gravitational potential will have the form $f_\phi G \Mvir / \rvir$, where
$f_\phi$ is a factor that depends on the exact density profile,
e.g.~$\sim 5$ for an NFW profile with a halo concentration of 5
\citep{Binney2008}.
For a virial overdensity $\Delta_c \simeq 200$ and absolute density
$\rho_\mathrm{crit}(z) \simeq \rho_\mathrm{M,0}(1+z)^3$ (valid at high $z$),
the virial mass and radius are related as
$\rvir = (3 / [4\pi \Delta_c \rho_\mathrm{M,0}])^{1/3} (1+z)^{-1} \Mvir^{1/3}$.

The accreted gas sparks star formation, and the star formation rate will thus
be $\mathrm{SFR} = \epsilon \dMdt / (1-R+\eta)$, where the proportionality
accounts for the accretion efficiency
$\epsilon = 0.5$--1 \citep{Bouche2010}, gas which is
quickly recycled from short-lived stars with a return fraction $R\sim0.3$--0.5
\citep{Vincenzo2016}, and gas ejected due to galactic outflows with typical
mass loading factors $\eta = \dot{M}_\mathrm{out} / \mathrm{SFR} \sim 0.5$--3
\citep{Schroetter2015, Heckman2015}. Expressing the accretion rate in this way
allows us to compare to the \lya\ caused by star formation.

Thus we may write
\begin{eqnarray}
\label{eq:Lcool}
\nonumber
L_{\mathrm{Ly}\alpha\mathrm{,cool}}
          \hspace{-2mm}& \sim \hspace{-2mm}& f_\gamma f_\alpha |\Delta\Phi| \dMdt\\
\nonumber
          \hspace{-2mm}& \sim \hspace{-2mm}& f_\gamma f_\alpha f_\phi \frac{G \Mvir}{\rvir} \frac{1-R+\eta}{\epsilon} \mathrm{SFR}\\
          \hspace{-2mm}& \sim \hspace{-2mm}& 0.5\text{--}2 \times 10^{41}
                                             \mathrm{erg}\,\mathrm{s}^{-1}
                                             \frac{1+z}{1+8.8}
                                             \frac{\mathrm{SFR}}{\Msun\,\mathrm{yr}^{-1}}
                                             \left( \frac{\Mvir}{\ten{11}\Msun} \right)^{2/3}.
\end{eqnarray}

Having expressed the accretion rate in terms of SFR, we can then compare to the
\lya\ originating from star formation \citep{Kennicutt1998}:
\begin{equation}
\label{eq:LHII}
L_{\mathrm{Ly}\alpha\mathrm{,SF}} \sim 10^{42}
  \frac{\mathrm{SFR}}{\Msun\,\mathrm{yr}^{-1}} \, \mathrm{erg}\,\mathrm{s}^{-1},
\end{equation}
and see that cooling radiation will account for roughly 10\% of the total
emitted \lya, also bearing in mind that in dusty galaxies,
$L_{\mathrm{Ly}\alpha\mathrm{,SF}}$ will be more susceptible to absorption than
$L_{\mathrm{Ly}\alpha\mathrm{,cool}}$, since the former is produced by stars
which tend to inhabit the same regions as the dust, while the latter
predominantly is produced farther out in the galaxy.
In reality, the mass loading factor decreases somewhat with mass
\citep[e.g.][]{Dutton2009,Hayward2016,Muratov2015} such that the dependency on
$\Mvir$ in \eq{Lcool} will be shallower for larger masses than the exponent
$2/3$ suggests. This result is consistent with \citet{Faucher-Giguere2010} who
find a fraction of $\sim0.18$ at \z\ and $\Mvir=\ten{11}\Msun$ through similar
considerations.

%end

\subsubsection{Intergalactic radiative transfer}
\label{sec:IGMRT}

Because of the high neutral fraction of the Universe at \z, in general the blue
peak of the \lya\ emission line is completely suppressed; as light leaves a
galaxy, the CGM and the IGM starts ``erasing'' the spectrum from around the
line centre, or even somewhat into the red wing, gradually working its way out
through the line towards bluer and bluer wavelengths. As the lines are
broadened substantially due to the scattering in the ISM, the necessary
distance travelled before the full line is out of resonance with the CGM is
longer than our computational box. The often-used approximation of simply
removing the blue half of the spectrum is rather imprecise and will
underestimate the impact of the IGM at these high redshifts where the density
of neutral gas is so high.
Even if IGM absorption were not an issue, a
neighbouring galaxy at a modest distance can still induce a sizable reduction
of both peaks.
%\footnote{For instance, for a LAE at \z\ an intervening galaxy
%with a column density of $\NHI=2\times10^{20}\pcms$ (a ``damped \lya\
%absorber'') even at a distance from the emitter of 5 comoving Mpc, may erase
%most of the blue and more than half of the red peak, depending somewhat on the
%width of the emitted line.}.

To achieve a less stochastic
description of the observed profiles we use another approach: ``Inside'' the
galaxies, where photons are all the time scattered both \emph{into} and
\emph{out of} the LOS, we use the accurate RT described in \sec{galRT}, but
after a certain distance $r_0$ from the centre, when the probability of being
scattered into the LOS can be neglected, we instead multiply the spectrum at
that point with the \emph{average transmission curve} $F(\lambda) =
\ave{e^{-\tau(\lambda)}}$.  Here, the average is taken over \ten{3} sightlines
per galaxy in a cosmological volume, and a confidence interval (CI) is defined
by the range of $F$ enclosing 68\% of the individually calculated transmission
functions. To calculate $F(\lambda)$, we use the publically available RT code
{\sc IGMtransfer}.
%\footnote{\href{http://www.dark-cosmology.dk/~pela/IGMtransfer.html} {\tt{www.dark-cosmology.dk/\~{}pela/IGMtransfer.html}}.}
{\sc IGMtransfer} has previously been used to model the impact of the CGM on
\lya\ line profiles at lower redshift; \citet{Laursen2011} found that setting
$r_0 \sim 1.5 \rvir$ was a reasonable value of the threshold between the full
and the approximative RT scheme. At \z, however, where the large neutral
fraction of the IGM gives rise to significant scattering out to large distances
from the galaxies, this value will underestimate the transmitted flux
considerably.
% In App.~\ref{app:r0} we show that at \z\ the full RT preferably should be
% calculated out to $\sim 10 \rvir$.
Hence, in this work we calculate the full RT out to $10 \rvir$, which
convergence tests reveal to be sufficient.

%end IGMRT

%end LyaRT

%end obs

\section{Results}
\label{sec:res}

\subsection{Stellar masses}
\label{sec:Ms}

\begin{figure}%[!t]
\centering
\includegraphics [width=0.50\textwidth] {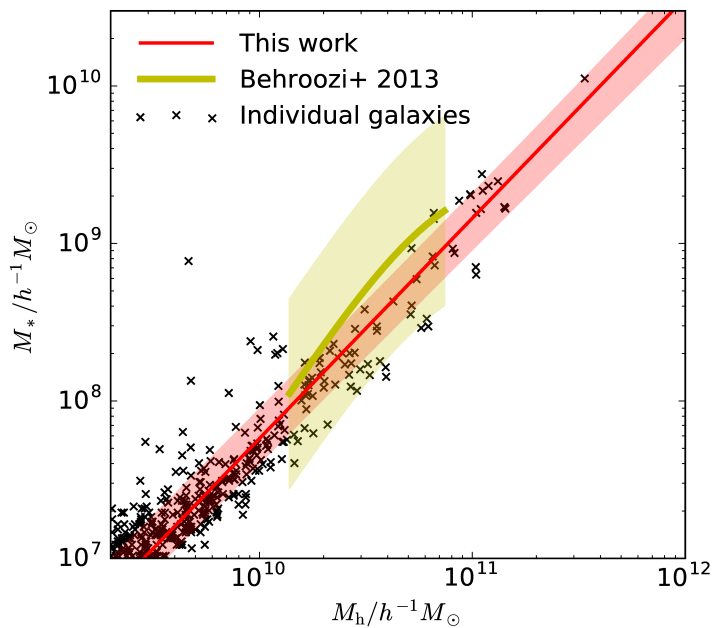}
\caption{Stellar mass $M_\star$ as a function of halo mass \Mh\ for individual
  galaxies in the hydro-simulation (\emph{black crosses}), and a power law fit
  (\emph{red}) to this distribution. For comparison, a model of the stellar
  mass-halo mass relation by \citet{Behroozi2013} at \z\ is shown in
  \emph{yellow}. Shaded areas show the 68\% CIs.}
\label{fig:M_star}
\end{figure}

\Fig{M_star} shows, as a function of halo mass, the stellar mass of the
simulated galaxies.
Fitting a power law to the distribution for masses above $\ten{9}M_\star$,
we find that the stellar mass is related to the virial mass roughly as
\begin{equation}
\label{eq:MsMh}
M_\star \sim 1.5\e{9} \left( \frac{M_\mathrm{h}}{\ten{11}} \right)^{1.4},
\end{equation}
where masses are measured in \hmMsun. \citet{Behroozi2013} present a
model of the relation between stellar masses and halo masses which is
consistent with observed stellar mass functions, specific SFRs, and cosmic SFRs
from $z=0$--8. Extrapolating the model to \z\ (in the \Mh\ range that
\citet{Behroozi2013} use for $z=8$), \fig{M_star} shows that our stellar masses
are in rough agreement with their model.
In the following figures, most quantities are shown as functions of halo mass,
because this is what is predicted from the HMF, but a secondary $y$ axis in the
top of the plots display an approximate scale for the corresponding stellar
mass, based on the relation in \eq{MsMh}.

\begin{figure}%[!t]
\centering
\includegraphics [width=0.50\textwidth] {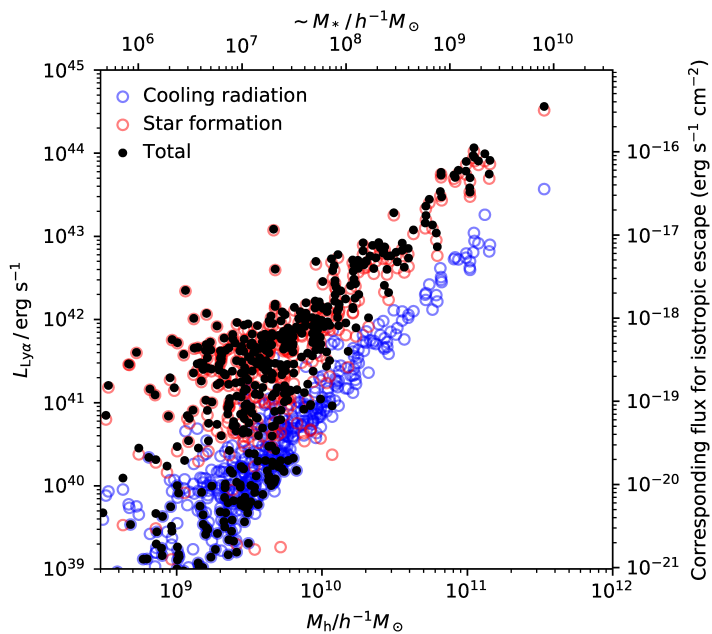}
\caption{Total, ``intrinsically'' emitted \lya\
  $L_{\mathrm{Ly}\alpha\mathrm{,tot}}$ (\emph{black dots}), and the
  contributions from star formation ($L_{\mathrm{Ly}\alpha\mathrm{,SF}}$;
  \emph{blue circles}) and cooling radiation
  ($L_{\mathrm{Ly}\alpha\mathrm{,cool}}$; \emph{red circles}). The secondary
  $y$ axis on the right shows the corresponding flux that \emph{would} be
  measured at Earth, if the \lya\ photons escaped isotropically, and were not
  absorbed by dust or scattered out of the LOS by neutral hydrogen. As seen in
  \fig{F_tot}, the actual measured values are quite a lot smaller.} 
\label{fig:L_int}
\end{figure}

As discussed in \sec{LyaRT}, a fraction of the stars emit enough ionizing
radiation to give rise to \lya\ emission; from the \ion{H}{ii} regions,
assuming a \citet{Salpeter1955} IMF, $1.1\e{42}\ergs$ is produced per SFR of
$1\,\Msun$ yr$^{-1}$ \citep{Kennicutt1998}, while for a \citet{Chabrier2003}
IMF the \lya\ luminosity is a factor $\sim1.8$ times higher, with a modest
dependency on stellar mass. In addition to this, accreting gas gets
shock-heated, subsequently cooling and emitting \lya. \Fig{L_int} show the
``intrinsically'' emitted \lya\ --- i.e.~before any RT effects --- for the two
processes.  
For masses up to $\Mh\sim\ten{10}\hmMsun$, star formation is sufficiently
stochastic that cooling radiation can dominate the total (albeit small)
luminosity, but at the high-mass end, the contribution is of the order of ten
per cent, in accord with our estimate in \sec{Lcool}.

%end Ms

\subsection{UV luminosity function}
\label{sec:UVLF}

As discussed in the introduction, the \lya\ luminosity is primarily linked to
star formation, specifically the massive O and B stars. Since these stars are
responsible not only for \lya\ but for the bulk of the entire UV spectrum, the
\lya\ LF and the UV LF are interlinked \citep[see, e.g.][]{Gronke2015a}, so
comparing the predicted UV LF to observations will support our confidence in
the simulations.  Because of the small size of our cosmological volume, it
cannot be used for determination of the HMF. Instead we use the ST HMF
(\eq{HMF}, calibrated to the Bolshoi simulation) to give us the number density
of halo masses, and then translate this to a LF by using the relation between
halo mass and brightness for our simulated galaxies. That is, 
%expressing the HMF as $[\mathrm{HMF}] = dN/d\log \Mh$,
with halo number density and
masses measured in $h^3\mathrm{Mpc}^{-3}$ and $h^{-1}\Msun$, respectively,
the LF can be written as
\begin{equation}
\label{eq:UVLF}
% \frac{\Phi}{\Mpc^{-3}\,\mathrm{mag}^{-1}} = h^3 [\mathrm{HMF}]
%   \left( \frac{d(M_\mathrm{UV}/\mathrm{mag})}{d\log(\Mh/\hmMsun)} \right)^{-1}.
\frac{\Phi}{\Mpc^{-3}\,\mathrm{mag}^{-1}} = h^3 \ln10 \,
 % [\mathrm{HMF}]
  \frac{dN}{d\ln\Mh} 
  \left( \frac{dM_\mathrm{UV}}{d\log\Mh} \right)^{-1}.
\end{equation}
%
% Note that $\Phi$'s volume is measured in $\Mps^{-3}$ rather than \phMpcc, to
% allow for direct comparison with observations.

In the following section, we describe how to obtain the last term in \eq{UVLF}.

\subsubsection{Relation between halo mass and UV luminosity}
\label{sec:MUV_Mh}

\citet{Trac2015} argue for a triple power law relation between the UV
luminosity and the halo masses. In terms of magnitudes this may be written
\begin{eqnarray}
\label{eq:MUV_Mh}
\nonumber
M_\mathrm{UV} & = & -2.5 \left[ \log L_0
                +     a         \log\left(    \frac{\Mh}{M_{\mathrm{h},a}} \right) \right. \\
              & + & (b-a)\left. \log\left(1 + \frac{\Mh}{M_{\mathrm{h},b}} \right) \right. \\
\nonumber
              & + & (c-b)\left. \log\left(1 + \frac{\Mh}{M_{\mathrm{h},c}} \right) \right] + 51.6
\end{eqnarray}
where $L_0$ is an overall amplitude, $M_{\mathrm{h},a}$, $M_{\mathrm{h},b}$, and
$M_{\mathrm{h},c}$ are three characteristic mass scales, and $a$, $b$, and $c$
are three power law slopes.

To determine the UV magnitudes of the simulated galaxies, each galaxy's UV
spectrum is calculated using the stellar population synthesis code STARBURST99
\citep{Leitherer1999}. Subsequently, for each galaxy \ten{4} sightlines are
followed from the locations of star formation in random directions, with the
probability of starting a sightline from a given location proportional to the
UV luminosity of stars in that region. Integrating dust extinction along the
lines of sight, a distribution of observed UV magnitudes at 1600 {\AA} in
different directions is obtained.

\begin{figure}%[!t]
\centering
\includegraphics [width=0.45\textwidth] {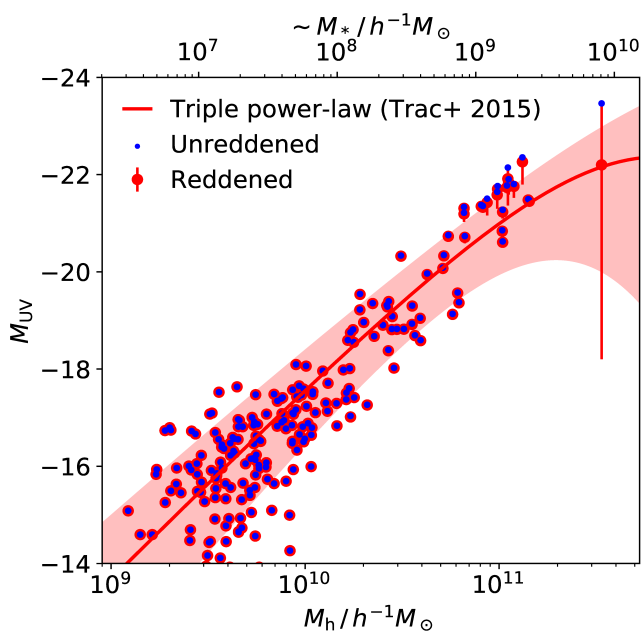}
\caption{UV magnitude $M_\mathrm{UV}$ at 1600 {\AA} as a function of their halo
         masses \Mh. For each galaxy the ``intrinsic'' UV magnitude calculated
         with STARBURST99 (\emph{blue dots}) and the reddened magnitudes
         calculated using the median and 68\% CI of the luminosity-weighted
         extinction along isotropically distributed sightlines from the star
         (\emph{red dots with errorbars}) are shown. The \emph{red line} shows
         the best estimate of the analytical relation in \eq{MUV_Mh}, with
         the \emph{red shaded} region showing the region within which 68\%
         measurements would fall.}
\label{fig:MUV}
\end{figure}

%new
\Fig{MUV} shows the halo mass--magnitude relation thus obtained, with the
reddened values (red points with error bars) showing the median and the 68\% CI
enclosed by the 16th and 84th percentiles. Comparing to the
\emph{un}reddened values (blue points), effects of reddening are seen to only
affect the most massive galaxies, with $M_\star\gtrsim\ten{9}\hmMsun$.

In addition to the simulated $M_\mathrm{UV}$ magnitudes, the best fit of the
functional form in \eq{MUV_Mh} with 68\% CI is shown in red. At low masses,
this CI primarily reflects a scatter in the galaxies' intrinsic luminosities,
while at high masses the anisotropic escape of the UV radiation due to dust, as
well as the fact that we have so few galaxies more massive than
$\sim\ten{11}\Msun$ comes into play.

%end MUV_Mh

\subsubsection{Comparing to observations}
\label{sec:compobsUV}

The derivative of \eq{MUV_Mh}, needed to evaluate \eq{UVLF}, is
\begin{equation}
\label{eq:dMUV_dMh}
\frac{dM_\mathrm{UV}}{d\log\Mh} = -2.5 \left(
                     a
                 + \frac{(b-a)}{1 + M_{\mathrm{h},b}/\Mh}
                 + \frac{(c-b)}{1 + M_{\mathrm{h},c}/\Mh} \right)
\end{equation}

\begin{figure}%[!t]
\centering
\includegraphics [width=0.45\textwidth] {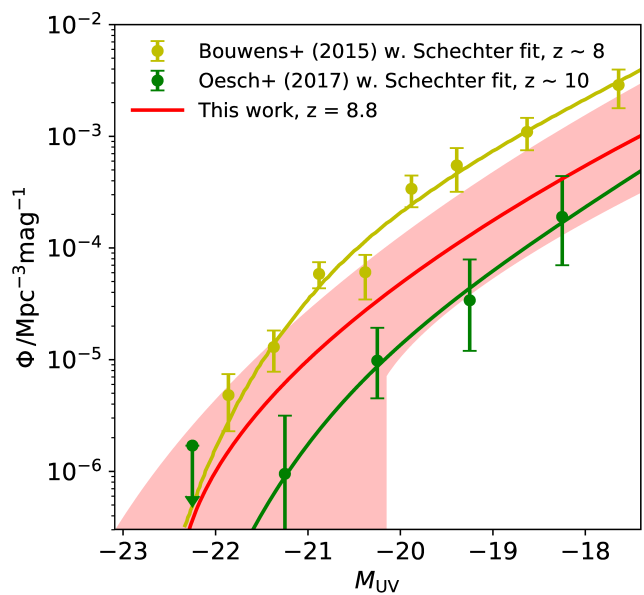}
\caption{Luminosity function $\Phi$ at 1600 {\AA} of the simulated galaxies
  (\emph{red solid}). The \emph{red shaded} region shows the 68\% spread in
  $\Phi$, which at the faint end reflects a scatter in the intrinsic UV
  luminosity, and at the bright end is due mostly to the anisotropic escape
  of UV radiation due to dust, as well as small-number statistics.
  \emph{Purple} and \emph{dark yellow} dots with errorbars show the observed
  LFs at $z\sim8$ \citep{Bouwens2015} and $z\sim10$ \citep{Oesch2017},
  respectively, with the associated \emph{solid lines} showing the best-fit
  \citet{Schechter1976}-functions.}
\label{fig:UVLF}
\end{figure}

The parameters of the fit to $M_\mathrm{UV}(\Mh)$ (\eq{MUV_Mh}) can now be
plugged into \eq{dMUV_dMh}, which in turn enables us to evaluate the UV LF
\eq{UVLF}. Due to the small number of galaxies detected at such high redshift,
no concensus of a $z\sim9$ LF exists, but \citet{Bouwens2015}
report a rather well-constrained UV LF at $z\sim8$, as well as a less
well-constrained one around $z\sim10$ which is improved in \citet{Oesch2017}.
\Fig{UVLF} shows our calculated UV LF, together with the those observations. It
is reassuring to see that the \z\ LF lies between those of $z\sim8$ and
$z\sim10$.
%

%end compobsUV

%end UVLF

\subsection{Typical line shapes}
\label{sec:lines}

The most prominent feature of a LAE is arguably the emission line profile;
much valuable knowledge can be extracted herefrom.
Scattering on neutral hydrogen slowly pushes the photons farther from the line
centre, so larger column densities are associated with broader emission lines.
The same is true for larger temperatures, as the thermal motion of hotter atoms
will engender larger Doppler shifts \citep[e.g.][]{Harrington1973,Neufeld1990}.
The relative height of the red and the blue peaks provides a measure of the
global kinematics of the galaxy, since blue (red) photons are suppressed by
outflowing (infalling) gas, by being Doppler-shifted into the frame of
reference of the gas elements \citep[e.g.][]{Dijkstra2006}.
At higher redshifts the blue peak is also influenced by the CGM
\citep[e.g.][]{Laursen2011}.
The longer the photons travel, the more prone they are to dust absorption,
so photons emitted in the dense and dusty star-forming regions, which would
have to scatter to the wings of the spectrum, escape less easily than photons
emitted in the outskirts of the galaxy (e.g.~from cooling radiation); hence,
dust is expected to narrow the spectrum \citep{Laursen2009b}.

The intertwinement of these effects, however, also makes the interpretation of
\lya\ observations notoriously difficult, motivating the need for a more
profound theoretical understanding.

% In general, the large cross section for photon at the line centre make the mean
% free path of such photons impossible. At each scattering, however, the motion
% of the atoms induces a small Doppler shift, eventually rendering the optical
% depth for the photon small enough to escape the galaxy. For an ISM with no
% collective large-scale motions, such as outflows or accretion, the emerging
% spectrum will thus be a double-peaked line

\begin{figure}%[!t]
\centering
\includegraphics [width=0.45\textwidth] {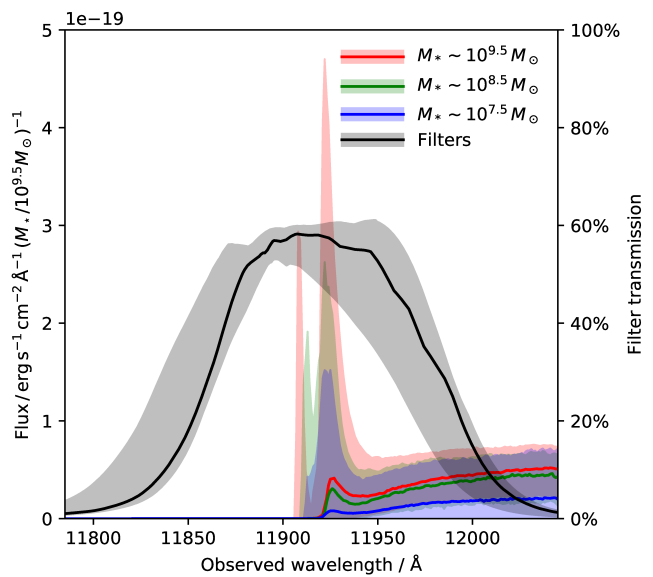}
\caption{Median spectra of galaxies in various halo mass ranges, normalized to
  the same stellar mass ($10^{9.5}\Msun$), and including the impact of the IGM
  on the line. The shaded areas shows the region in which 68\% per cent of the
  lines fall. Also shown, in \emph{black}, is the median transmission (right
  $y$ axis) of the 16 filters, with the grey region showing the 68\% CI. 
  Although the average flux in the line hardly rises above the continuum,
  prominent lines are seen for a significant fraction of the galaxies.}
\label{fig:spec}
\end{figure}

\Fig{spec} shows the average spectra for galaxies in different mass ranges.
In the figure, the spectra have been normalized to the same stellar mass,
namely $M_\star=\ten{9.5}\hmMsun$, which is the mass of the
most massive mass range. In each mass range a few tens of galaxies have been
used, and for each galaxy, the median and the dispersion of six directions,
each with 100 different realizations of the impact of the IGM, has been used.
While the average spectrum of even the most massive galaxies is virtually
erased by the IGM, within $1\sigma$ many lines do rise above the continuum.
Despite the normalization, the lines decrease in intensity with decreasing
stellar mass. While this may in part be attributed to less cooling
radiation, the main reason is that more massive galaxies are more likely to
ionize larger bubbles around themselves, facilitating the transfer of \lya.

%end lines

% \subsection{Surface brightness}
% \label{sec:SB}
% 
% %end SB

\subsection{Escape fractions}
\label{sec:fesc}

\begin{figure}%[!t]
\centering
\includegraphics [width=0.45\textwidth] {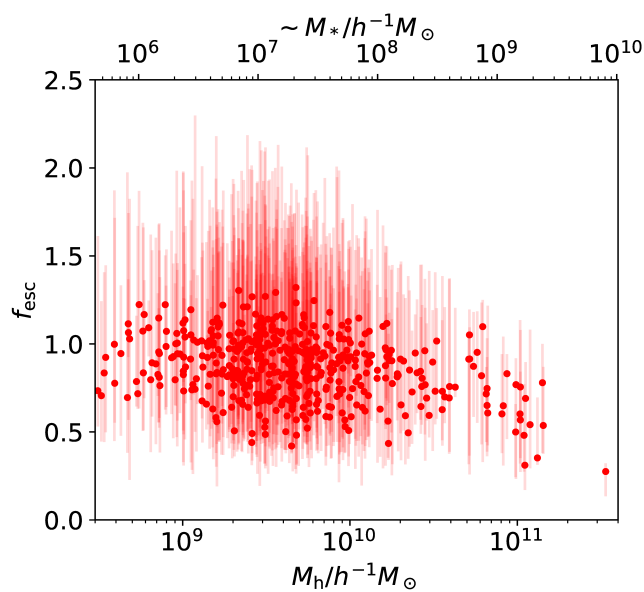}
\caption{Escape fraction \fesc\ of \lya\ as a function of halo mass \Mh,
  defined as the ratio of the number of \lya\ photons that make it out through
  the galaxy to the total number emitted. The points and the error bars denote
  the median and the 68\% CI of the six directions along the Cartesian axes.
  If there were no dust and galaxies were isotropic, \fesc\ would be 1. The
  extent of the error bars is chiefly dictated by the anisotropy of the ISM.
  Only for galaxies above $\Mh\sim\ten{10}\hmMsun$ does dust play a significant
  role for the escape fraction. After escaping the galaxy, photons may still be
  scattered out of the LOS, so the observed \fesc\ would be (much) smaller
  than these values.} 
\label{fig:f_esc}
\end{figure}

As soon as the first generation of stars end their lives, the ISM is polluted
with metals. A fraction of these metals deplete to form dust grains.
% There has
% been some debate on whether the dust-to-metal ratio $f_{\mathrm{dm}}$ was
% smaller in the past, such that dust builds up only slowly. Observations of
% $f_{\mathrm{dm}}$ at high redshifts are sparse, but to date, there seems to be
% no evidence that it should be substantially different from the local value of
% $\sim 1/2$ \citep[e.g.][]{Pettini1997,Savaglio2003}. In fact, theoretical
% models of the evolution of dust predict that it should build up over 10--100
% Myr after the first metals emerge \citep{Gall2011}, while observations of
% absorption lines in background QSO and GRB spectra indicate that the formation
% time scales are but a few Myr \citep{Zafar2013}.
As galaxies must probably have gone through at least some star formation
episodes to make themselves visible, it therefore comes as no surprise that
most visible galaxies must also contain some amount of dust. Because of the
long path length of \lya\ (compared to continuum photons) due to scattering,
\lya\ is particularly prone to dust absorption\footnote{Note that,
theoretically, it is possible to have the reverse effect, i.e.~to have a dusty,
multiphase ISM that preferentially lets \lya\ escape
\citep{Neufeld1991,Hansen2006}. However, as was shown by \citet{Laursen2013}
and \citet{Duval2014} this requires physically unrealistic scenarios; in
particular an almost vanishing velocity field together with (super-)Solar
metallicities and very high density contrasts between the phases.}.
Observationally, it is difficult to distinguish between photons lost to dust
and photons scattered out of the LOS by the IGM, but in the simulations this is
possible.  However,
%as shown in \fig{EBV},
only the most massive galaxies have been able to
generate enough dust to give to a measurable colour excess; 
thus, averaged over all directions the \lya\ escape fraction \fesc\ is close to
unity. Nevertheless, the complex density field entails a quite anisotropic
escape, where some sightlines are blocked by dense clouds, instead beaming the
radiation in other directions. Defining the escape fraction as the ratio of
observed \lya\ flux to what would be observed if all photons escaped
isotropically, \fesc\ will thus exceed unity in some directions. This is seen
in \fig{f_esc}, where also the large spread between different sightlines is
noticed.  
%
% In general, however, although a surplus of \lya\ is beamed in certain
% directions, the large neutral fraction of hydrogen scatters away so much
% radiation that the observed \fesc\ is below unity.

%end fesc

\subsection{Observed flux}
\label{sec:flux}

\subsubsection{Optimal aperture}
\label{sec:ap}

As discussed in \sec{IGMRT}, scattering in the IGM persists out to many virial
radii. Photons travelling in the direction towards the observer are lost from
the LOS, but do not vanish; instead they become part of the background. 
Similarly, photons that leave the galaxy initially in another direction, have a
small chance of being scattered towards the observer. This mechanism creates a
halo of \lya\ light around the galaxy, which has indeed been observed at all
redshifts down to $z\sim0$ \citep{Hayes2013,Leclercq2017}.

\begin{figure}%[!t]
\centering
\includegraphics [width=0.40\textwidth] {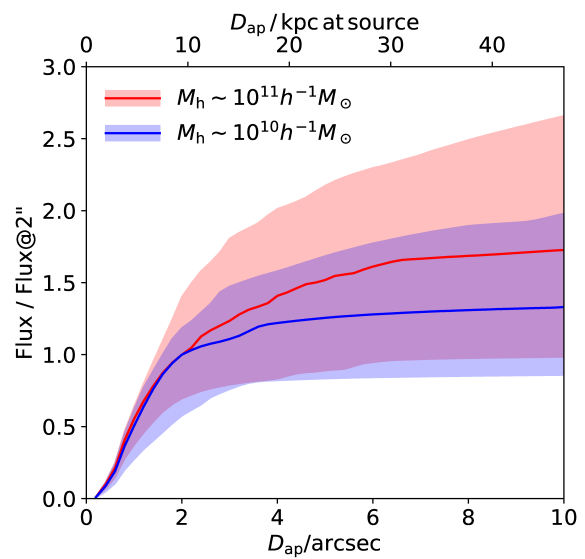}
\caption{Measured flux as a function of aperture diameter $D_{\mathrm{ap}}$,
  normalized to the flux measured at $D_{\mathrm{ap}} = 2''$, for
  $\Mh\sim\ten{11}M_\odot$ (\emph{red}) and
  $\Mh\sim\ten{10}M_\odot$ (\emph{blue}) galaxies, with the shaded regions
  giving the 68\% confidence interval.
  For the $\Mh\sim\ten{11}M_\odot$ galaxies, the observed flux for
  $D_{\mathrm{ap}} = 2''$ is $\sim50$\% compared to
  $D_{\mathrm{ap}} \rightarrow \infty$, but since a smaller $D_{\mathrm{ap}}$
  means less noise, slightly better results are achieved for
  $D_{\mathrm{ap}} = 1\farcs4$.}
\label{fig:F_ap}
\end{figure}

The half-light radii of our simulated galaxies increase from a few kpc, to
roughly 5 kpc, to $\sim$10 or even tens of kpc for small, intermediate, and
massive halos, respectively.
When measuring the total flux received from a galaxy, a larger aperture will
always result in more flux, but of course also in more background noise.
The standard aperture for measuring flux has a diameter of $2''$, corresponding
at \z\ to just under 10 kpc. For the most massive galaxies, which as yet are
the only ones we have a chance of detecting, we will hence miss of the order
half of flux (see \fig{F_ap}).
In fact, due to the increased noise for large apertures, the ``optimal''
aperture turns out to be a slightly smaller 1\farcs4, although the
difference in the total number of observed galaxies is on the $<10\%$ level
when comparing to using a $2''$ aperture.

\citet{Sadoun2016} similarly discuss the fraction of photons that fall outside
an aperture of diameter $2''$ for a strong and a weak ionizing background. For
the weak UVB, corresponding to early in the EoR, they predict that the flux
should be suppressed by a larger factor of $\sim10$. However, their LAE model
is modeled as a smooth gas distribution, which forces photons to scatter to
larger distances than in a clumpy medium before being able to escape.

%end ap

%end flux

\subsection{Translating flux threshold to minimum observable mass}
\label{sec:F2M}

\begin{figure}%[!t]
\centering
\hspace*{-1cm}                                                           
\includegraphics [width=0.48\textwidth] {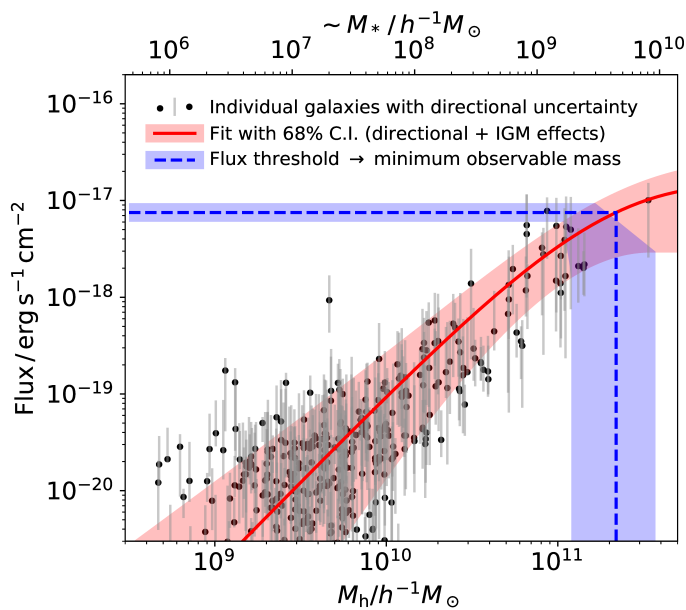}
\caption{Flux $F$ measured inside a 1\farcs4 circular aperture (\emph{black
  dots}) with errors given by the 68\% CI of the six directions of
  observation, as a function of halo mass \Mh. Additional to this ``intrinsic''
  variation, the inhomogeneous IGM, as well as the fact that galaxies of
  similar masses exhibit a variety in physical properties, introduces a scatter
  in possible values of observed flux; the 68\% CI of this scatter is
  represented by the \emph{red, shaded} region, while the \emph{solid, red}
  line is the double power law given in \eq{FMh}. The \emph{dashed, blue} line
  shows how the flux threshold of the detectors results in a minimum,
  observable halo mass $M_\mathrm{h,min}$, and the \emph{shaded, blue} region shows how the spread
  in detector thresholds, combined with the spread in fluxes, results in a
  spread in $M_\mathrm{h,min}$.}
\label{fig:F_tot}
\end{figure}

\Fig{F_tot} shows the expected observed flux $F$ measured within a 1\farcs4
circular aperture centered on the brightest pixel for the simulated galaxies as
a function of their halo mass. Viewing the galaxies from different directions
introduce a large scatter in the observed flux, both due to the anisotropy of
the galaxies, and due to the inhomogeneous IGM. The scattered points show the
median observed flux (i.e.~of the radiation that makes it through both ISM and
IGM), while the error bars show the 68\% CI in the flux introduced by the
anisotropic ISM alone.
Assuming a power law-like relation between $F$ and \Mh\
(which ought to be suitable at least over a limited mass range),
but with a turnover at high masses
where dust and possibly a more compact ISM makes \lya\ escape harder
\citep[see][]{Laursen2009b}, a double power law given by
\begin{equation}
\label{eq:FMh}
\frac{F(\Mh)}{\ergscm2\,\text{\AA}^{-1}} = \ten{-19}
  \left(     \frac{\Mh}{\ten{10}} \right)^{1.8}
  \left( 1 + \frac{\Mh}{ 3\e{11}} \right)^{-2.3}
\end{equation}
%%%%%%%%
%
%%%%%%%%%%%%%%%%%%%%%%%%%%%%%%%%%%%%%%%%%%%%%%%%%%%%%%%%
% % With units, divided on two lines
% \begin{eqnarray}
% \label{eq:FMh}
% \nonumber
% \frac{F(\Mh)}{\ergscm2\,\text{\AA}^{-1}} & = & \ten{-19}
%              \left(     \frac{\Mh}{\ten{10}\hmMsun} \right)^{1.8} \\
% & \times &   \left( 1 + \frac{\Mh}{ 3\e{11}\hmMsun} \right)^{-2.3}
% \end{eqnarray}
%%%%%%%%%%%%%%%%%%%%%%%%%%%%%%%%%%%%%%%%%%%%%%%%%%%%%%%%
with masses, again, measured in \hmMsun\ is seen to be a good fit to the
data in the relevant range (the negative exponent in the last term probably
makes this functional form impractical at higher masses).
The red, shaded region captures the 68\% CI from both the anisotropic escape,
the inhomogeneous IGM, and the spread in flux from galaxies of similar masses;
that is, it shows the most probable region that a new measurement would fall.
%
% the best fit to $F(\Mh)$ is shown as a red line in the plot, and is given by
% % %
% % \begin{equation}
% % \label{eq:FMh}
% % \frac{F(\Mh)}{1.8\e{-18}\ergscm2} = \left( \frac{\Mh}{\ten{11}\hmMsun} \right).
% % \end{equation}
% % %
% %
% \begin{equation}
% \label{eq:FMh}
% F(\Mh) = 1.8\e{-18} \left( \frac{\Mh}{\ten{11}\hmMsun} \right)^{1.7} \ergscm2.
% \end{equation}
% %
%%%%%%%%%%%%%%%%%%%%%%%%%%%%
% The shaded red region shows the 68\% CI, i.e.~the
% expected region inside which a measurement of a new galaxy would fall with 68\%
% certainty, including the spread in flux imposed by the IGM.
%%%%%%%%%%%%%%%%%%%%%%%%%%%%
% The fit and the CI was calculated by binning the data in 20 bins, computing
% the median and the 16th and 84th percentiles in each bin, and finally fit a
% power law to the median, lower, and upper individually in log-log space through
% standard, least-square fitting procedures. This approach was chosen in order
% to simultaneously wallah wallah wallah... 

% The dependence of the flux on halo mass is not straightforward, but may be
% expected on average to scale approximately according to some power law; the
% observed flux is a convolution of the size of the galaxy and its ability to
% form massive stars (and accrete gas), but also on its ability to form dust and
% the resulting effect on the escape fraction of \lya\ photons, all of which to
% some extend behaves in a self-similar fashion. Assuming a power law scaling
% between the halo mass and the observed flux, a relation

The 16 detectors of \uv\ have a varying threshold for detecting sources at a
given significance, and even have variations in different regions of the
detectors. For a 5$\sigma$ detection and an aperture of 1\farcs4, they have an
average threshold of $m_\mathrm{AB,thres} = 25.25\pm0.24$,
%. For an average FWHM
%of $123\pm3$ {\AA} and central wavelength of $11\,914\pm16$ {\AA}, this
%corresponds to a flux threshold of
corresponding to a flux threshold of
$F_\mathrm{thres} = (7.5_{-1.9}^{+1.5})\e{-18}\ergscm2$,
shown in \fig{F_tot} as the horizontal, blue, dashed line with the shaded
region marking the spread. These values are as measured in a 1\farcs4 aperture,
and do not include an aperture correction, since we measure the flux in our
simulated images using this aperture.

The first thing to notice is that indeed some of the simulated galaxies make it
above the flux threshold, at least in certain directions --- the question is
whether such galaxies are sufficiently common that there is a significant
probability that \uv\ will detect them.  Assuming the relation \eq{FMh} to hold
true, this flux threshold can be translated into a \emph{minimum
observable halo mass}
\begin{equation}
\label{eq:Mmin}
M_\mathrm{h,min} = (2.2_{-1.0}^{+1.5})\e{11}\hmMsun,
\end{equation}
seen in \fig{F_tot} as the vertical, blue, dashed line. This corresponds
roughly to a minimum observable \emph{stellar} mass of
$M_{\star\mathrm{,min}} = (5_{-3}^{+5})\e{9}\hmMsun$.

%end F2M

\subsection{Translating minimum observable halo mass to expected number of
  observed galaxies}
\label{sec:M2N}

\begin{figure}%[!t]
\centering
\includegraphics [width=0.45\textwidth] {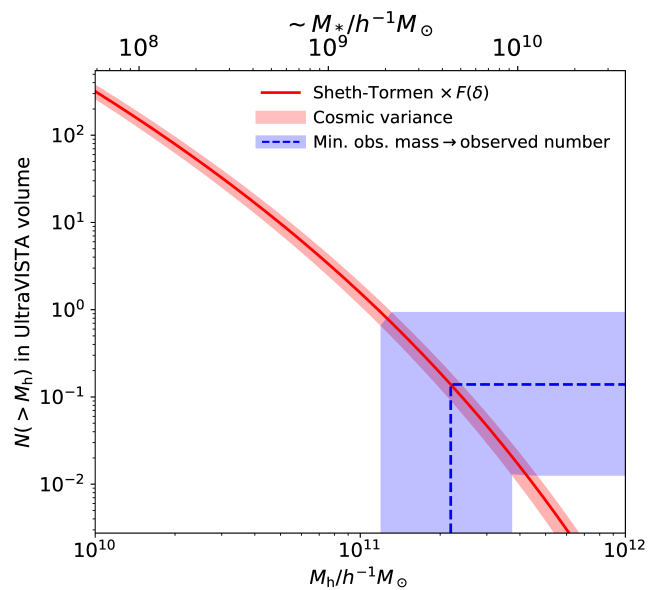}
\caption{Cumulative halo mass function (\emph{solid red}), normalized to
  the approximate volume surveyed by \uv. The \emph{red-shaded} region
  represent the 1$\sigma$ standard deviation on the number counts resulting
  from cosmic variance. The \emph{dashed blue} line shows the minimum
  observable halo mass (\eq{Mmin}) in the \uv\ survey and the corresponding
  number of observed galaxies. The \emph{blue-shaded} region shows the 68\%
  confidence interval for this.}
\label{fig:cumMF}
\end{figure}

Integrating the HMF from a given mass scale $M'$ to infinity, the total number
of haloes at least as massive as $M'$ is found. \Fig{cumMF} shows the resulting
\emph{cumulative halo mass function} in the volume surveyed by \uv.

\subsubsection{Cosmic variance}
\label{sec:cosvar}

Density fluctuations in the large-scale structure lead to uncertainties in the
number counts, known as \emph{cosmic variance}.
The expected standard deviation due to this effect is shaded red in \fig{cumMF}.
This uncertainty is obtained following \citet{Trenti2008}, who estimate the
cosmic variance employing a combination of excursion set theory
\citep{Bond1991} and $N$-body simulations. We have made use of the associated
online
calculator\footnote{\href{http://casa.colorado.edu/~trenti/CosmicVariance.html}
{\tt{casa.colorado.edu/\~{}trenti/CosmicVariance.html}}.} which allows for a
\citet{Sheth1999} bias instead of a \citet{Press1974} bias, as well as changing
the value of $\sigma_8$ to our preferred value of 0.82.
%instead of their value of $\sigma_8 = 0.75$ and 0.9.
All their simulations have
$\{\OmegaM,\OmegaL\} = \{0.26,0.74$\}, rather close to the
$\{0.27,0.73\}$ of the Bolshoi simulation. Taking into account cosmic variance
results in a relative uncertainty on the number of galaxies increasing from
17\% at $\Mh = \ten{10}\hmMsun$ to 48\% at $\Mh = \ten{12}\hmMsun$.
% At $M_\mathrm{h,min} = (2.3_{-0.8}^{+1.1})\e{11}\hmMsun$, the uncertainty is
% $33\pm3$\%.

The vertical blue dashed line shows the minimum observable halo mass
$M_\mathrm{h,min}$, discussed in the previous section, and the horizontal line
shows the corresponding number of galaxies that will be observed.
When the 68\% CI of the $M_\mathrm{h,min}$, shown as the blue-shaded region,
is combined with the cosmic variance, the resulting expectation value of the
number of observed galaxies is $N(\Mh \ge M_\mathrm{h,min}) =
0.12_{-0.11}^{+0.79}$.  

%end cosvar

\subsubsection{Poisson noise}
\label{sec:poisson}

In addition to cosmic variance, there will be an uncertainty from usual Poisson
noise. For large values of $N$, the Poisson noise converges towards $\sqrt{N}$,
but for small values this approximation becomes increasingly inaccurate, as the
68\% CI becomes increasingly asymmetric\footnote{For instance, the Poisson
uncertainty on the numbers 100 and 10, is not 10 and 3.16 as one might naively
expect, but $_{-9.98}^{+11.0}$ and $_{-3.11}^{+4.27}$, respectively
\citep{Gehrels1986}.}.

Since the number of galaxies is an integer, it makes little sense to quote a
number such as the one found in the previous section; instead it is more
illuminating to find the probability of detecting a given integer number.
From the standard definition of the Poisson distribution, we can then finally
state the disheartening probabilities of observing 0, 1, or 2 or more galaxies
as
\begin{eqnarray}
\label{eq:P_N}
\nonumber
P(0)  & = & 88_{-45}^{+11} \%, \\
P(1)  & = & 11_{-10}^{+25} \%, \\
\nonumber
P(2+) & = &  1_{- 1}^{+14} \%.
\end{eqnarray}
%

%end poisson

%end M2N

\subsection{Luminosity function}
\label{sec:LF}

\begin{figure}%[!t]
\centering
\includegraphics [width=0.45\textwidth] {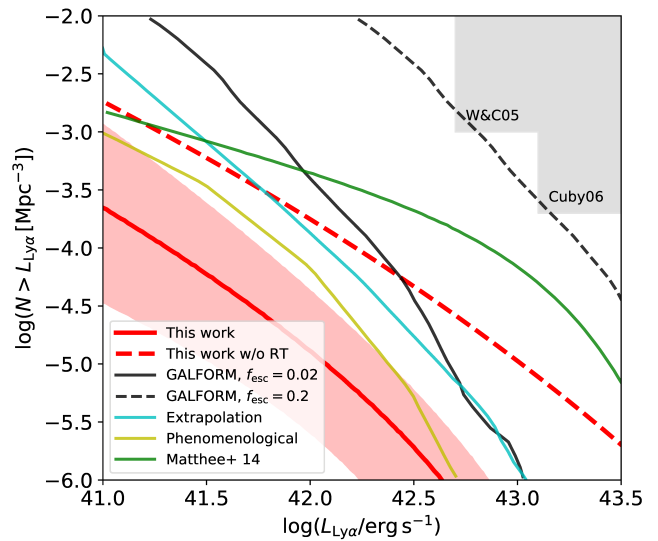}
\caption{\lya\ LF predicted from our simulations (\emph{red} with shaded area
         showing the 68\% CI), compared to the three models of
         \citet[][described in \sec{uv}]{Nilsson2007}. Also shown, with the
         \emph{red, dashed} line, is the \lya\ LF predicted from our
         simulations if we did not take into account RT effects.
         Additionally, the ``optimistic'' LF of \citet{Matthee2014} is shown
         in \emph{green}.
         The \emph{grey} region marks observational upper limits from
         \citet{Willis2005} and \citet{Cuby2007}.} 
\label{fig:LyaLF}
\end{figure}

Following the same approach as in \sec{UVLF}, a \lya\ LF can be constructed.
This is seen in \fig{LyaLF}, where we also compare to the models presented in
\citet{Nilsson2007} and \citet{Matthee2014}; the latter is the most optimistic
LF allowed by the limits of their survey.
Our LF lies well below all models, but is less steep at the bright end,
compared to the closest models.

\Fig{LyaLF} also shows how the predicted \lya\ LF would look if we did not take
into account \lya\ RT. At the bright end, this differs from the correct LF by
almost two orders of magnitude, highlighting the crucial role of a proper RT
treatment.

%end LF

%end res

\section{Discussion}
\label{sec:disc}

\subsection{Colour selection --- differing impact of the IGM on narrowband and
            broadband}
\label{sec:Pselect}

In \sec{M2N} we predicted that the probability of detecting just one, or
two or more, galaxies in the \uv\ survey is of the order of 10\% and 1\%,
respectively. But even if a galaxy is brighter than the detection threshold, it
still needs to be selected as a narrowband excess object,
 i.e. to have a higher flux density in a
narrowband than in a broadband in the same wavelength region, in this case the
$J$ band.

If there were no emission line, a flat spectrum would have $\mathrm{NB} -
\mathrm{BB} = 0$, where NB is the magnitude in the \lya\
narrowband filter, and BB is the $J$ band magnitude.
However, at
such high redshifts where the IGM suppresses not only the blue half of the
\lya\ line, but also a significant part of the red half, the NB and the BB are
affected differently. The reason is that the NBs are centered on the line, so
that (less than) 50\% of the flux is transmitted, whereas in the $J$ band ---
which transmits in $\lambda \sim [11\,650\text{--}13\,500]$ {\AA} the \lya\
break is located quite far to the blue,
% As a rough estimate, in order to obtain
% a more reasonable selection criterion, let us assume that 50\% and 85\% of the
% flux in the NB and BB is transmitted, respectively. In that case, a flat
so that roughly 85\% is transmitted. For 50\% transmission in the NB, a flat
spectrum with no emission line would have a zeropoint of $\mathrm{NB} -
\mathrm{BB} = -2.5\log(50/85)\simeq0.6$. In reality, the NB transmission is
closer to 25\%, in which case the zeropoint lies at $\sim1.3$.
In these calculations, the \lya\ line centers have been assumed to coincide
with the NB center.

\begin{figure}%[!t]
\centering
\includegraphics [width=0.45\textwidth] {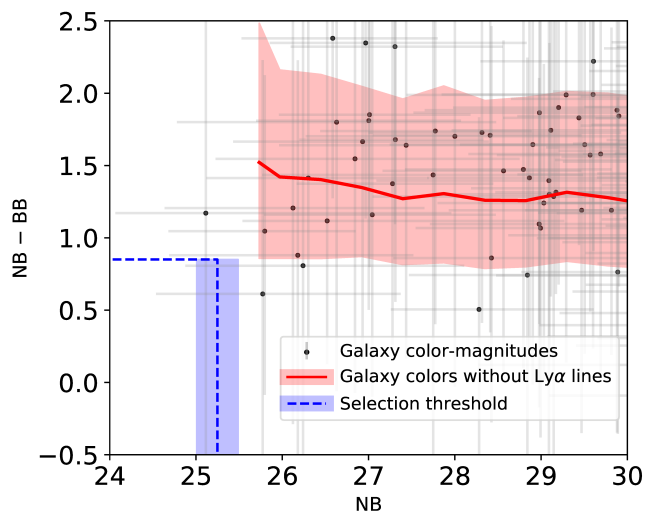}
\caption{Colour-magnitude diagram of the simulated galaxies. \emph{Black dots}
  show individual galaxies, with \emph{grey} error bars showing the 68\% CI
  for different sightlines. The \emph{red} line shows a running median (with
  shaded 68\% CI) of the $\mathrm{NB} - \mathrm{BB}$ colour as a function of
  NB magnitude for all galaxies in the hypothetical case where they would have
  no \lya\ emission lines. The selection criteria are shown with \emph{blue
  dashed} lines; the magnitude threshold is the same as in \ref{sec:F2M}, while
  the colour threshold is set to $\mathrm{NB} - \mathrm{BB} < +0.85$,
  corresponding roughly to the lower limit of the ``no \lya''. Although no
  galaxy on average falls inside the selection square, the directional variance
  results in a probability of a few tens of \% of being selected, albeit only
  for the brightest galaxies.}
\label{fig:NB-BB}
\end{figure}

\Fig{NB-BB} shows a colour-magnitude diagram of the simulated galaxies.
The error bars correspond to different LOSs. To calculate a more realistic
zeropoint, the colours and magnitudes are also calculated in the hypothetical
case where the galaxies have no \lya\ emission lines. In this case, the scatter
is much smaller and is caused mainly by IGM inhomogeneities in the IGM.
A running median (with 16th and 84th percentiles) of the galaxies' colours in
this case is shown in the figure in red. The median colour is seen to be rather
independent of NB magnitude, with a value around $\mathrm{NB} - \mathrm{BB}
\simeq 1.25$, except for the brightest galaxies where dust begins to play a
role; here the reddening causes a slight upturn in the colour index. The lower
limit of the 68\% CI lies around $\mathrm{NB} - \mathrm{BB} \simeq +0.85$.
Using this value as the colour threshold, together with the magnitude limit of
the individual detectors, a ``selection square'' is defined, shown with blue,
dashed lines in \fig{NB-BB} (the blue-shaded region shows the spread in the
detector limits).

Considering only the median colours and magnitudes of the galaxies, not even the
brightest ones fall inside the selection square. However, taking into account
the variability in brightness in different directions, the brightest ones do
in fact have a non-zero probability $P_\mathrm{sel}$ of being selected as LAEs.
The brightest galaxy in the sample, which is also the most massive with
$\Mh=3.4\e{11}\hmMsun$, has a 35\% chance of being selected, i.e.~to meet the
colour selection criterion \emph{and} being brighter than the detection limit
of the survey. There is no clear
trend of $P_\mathrm{sel}$ with halo mass, but for
$\Mh\gtrsim\ten{11}\hmMsun$, $P_\mathrm{sel}$ lies between a few \% and 35\%,
while for $\Mh\lesssim\ten{11}\hmMsun$, $P_\mathrm{sel}$ is virtually zero.

%end Pselect

\subsection{Halo mass function and cosmology}
\label{sec:HMFdisc}

The HMF was, and typically is, calculated considering dark matter only.
Although baryons make up only $\sim1/6$ of the total mass, their ability to
cool and condense leads to more concentrated mass profiles, resulting in
principle in higher halo masses. Thus, \citet{Stanek2009} and \citet{Cui2012}
find an increased number density of cluster scale haloes in simulations
including gas cooling and star formation. However, also taking into account the
effect of AGN feedback prevents overcooling of the gas, counteracting this
concentration.  \citet{Martizzi2014} find that, incidentally, the resulting
mass function is quite close to the DM-only HMF. We therefore choose to ignore
baryonic effects on the HMF.

Of greater concern might be the functional form of the multiplicity function
$f(\sigma)$ (\eq{fsigST}). Both the PS and the ST HMFs are ``universal'', in
the sense that their multiplicity functions are independent of power spectrum
and expansion history of the Universe.
Several successive authors have noted departures from universality
\citep[e.g.][]{Tinker2008,Crocce2010,Watson2013}.
\citet{Tinker2008} let the parameters of $f(\sigma)$ be a function of
redshift, and in the BolshoiP simulation (an ``updated'' version of the Bolshoi
simulation with \citet{PlanckCollaboration2016} cosmology), \citet{Klypin2016}
showed that this HMF fitted their halo abundances well, albeit only in the
range $0<z<2.5$.

\citet{Behroozi2013} extended the \citet{Tinker2008} HMF to $z=8$,
%i.e.~close to the redshift of eventual \uv\ LAEs,
although only for masses up to 
$\Mh\sim\ten{11.5}\hmMsun$, which is lower than the upper limit for the
minimum detectable mass in \uv. However, \citet{Rodriguez-Puebla2016} provide
a fitting formula for a HMF that gives good fits to the BolshoiP simulation
at higher masses. Extending that fit to \z\ increases the number density of
haloes with masses around the minimum detectable halo mass in by a factor of
five from the $0.12_{-0.11}^{+0.79}$ found in \sec{cosvar} to $N(\ge\Mh) =
0.6_{-0.6}^{+4.2}$ or, in terms of probabilities of detecting an integer number
of galaxies,
$\{P(0), P(1), P(2+)\} = \{55_{-54}^{+42}, 33_{-29}^{+29}, 12_{-2}^{+21}\}$\%.
That is, although this still implies that a non-detection is the most likely
outcome, the prospects of finding a few are less dire.

%end HMFdisc

\subsection{Field galaxies vs. cluster galaxies}
\label{sec:fieldVsCluster}

The fact that the galaxies are taken from a proto-cluster region might be a
concern, since they could evolve differently from a ``typical'' galaxy.
However, although cluster galaxies form earlier, and peak earlier in star
formation, than field galaxies, the two have comparable specific SFRs
\citep{Muldrew2017}. Since our model assigns a brightness to a galaxy based on
its mass, but the mass distribution itself is taken from the HMF which is
unrelated to the simulated galaxies, we do not consider this to be an issue.

%end fieldVsCluster

\subsection{Ionization state of the IGM}
\label{sec:IGMdisc}

The largest uncertainty in our forecast is arguably the ionization state of
the IGM. Although the UV RT is treated accurately, it hinges on the assumed
evolution of the UVB field. The global neutral fraction in our hydro-simulation
is $\xHI\simeq0.13$, but in the IGM the fraction is much lower, of the order
$\ten{-4}$ to $\ten{-3}$.
This is still enough to erase the blue part of the
\lya\ line, and the CGM and residual neutral clumps cause the damping wing to
erase a large part of the red wing as well.

Even if we have overestimated the impact of the CGM/IGM, it is hard to imagine
that even a small fraction of the blue wing escape; although some galaxies are
surrounded by highly ionized bubbles,
galaxy and quasar spectra show complete Gunn-Peterson
troughs (e.g.~\citet{Fan2006}; see however \citet{Hu2016} and
\citet{Matthee2018} for a counterexample at $z=6.6$)
If we redo our analysis simply removing the
blue half of the spectrum --- thus neglecting the effect of the damping wing
--- the expected number of detected LAEs increases to
$1.0_{-1.0}^{+2.9}$. Only if we completely neglect the IGM do we predict a
large success rate, with $N = 4.5_{-4.1}^{+10.8}$.

%end IGMdisc

\subsection{Starburst-driven outflows}
\label{sec:outflows}

When massive stars explode as supernovae, they heat up the ISM, causing it to
expand and form bubbles. If the SN explosion rate is sufficiently high, the SN
remnants overlap before they can cool radiatively, in which case these bubbles
--- assisted by super-Eddington photon pressure --- may extend beyond galactic
scales, manifesting themselves as outflowing superwinds \citep[see e.g.][and
references therein]{Powell2011}.

Such winds have been observed at lower redshifts, mainly through blueshifted
absorption lines \citep[e.g.][]{Kunth1998,Pettini2001,Shapley2003}, and are
usually thought to be the explanation for the often-observed asymmetric \lya\
line profiles \citep[e.g.][]{Verhamme2006,Tapken2007,Yamada2012}, where the red
peak is enhanced with respect to the blue.  One of the largest uncertainties in
hydrosimulations in general is arguably how the feedback from stellar energy
output affects the surrounding gas.  Although many cosmological simulations
have employed various implementations of feedback that are able to generate
outflows, when it comes to \lya\ RT in general the predicted line profiles tend
to be rather symmetric or even have an increased \emph{blue} peak
\citep[e.g.][]{Laursen2009a,Barnes2011,Yajima2012}.
That is, despite gas being ejected from
the galaxies, infalling gas accreting onto the galaxies dominate the
kinematics, leading to the opposite effect. Only in simulations of isolated
disc galaxies has a prominent red peak been successfully reproduced, and only
when observed face-on where the wind is strongest \citep{Verhamme2012}.

Observations based primarily on \ion{O}{vi} absorption lines
indicate that the CGM around star-forming galaxies at $z\sim0$ contains
considerable amounts of oxygen \citep[e.g.][]{Tumlinson2011,Prochaska2011}.
Also lower-ionization species such as \ion{Si}{ii}, \ion{Si}{iii}, and
\ion{N}{ii} are detected \citep[e.g.][and references therein]{Prochaska2017}.
Based on observations of potential \ion{O}{vii} absorption in the hot CGM of
the Milky Way, \citet{Gupta2012} infer the presence of large amounts of oxygen
in the CGM as well \citep[see however][]{Wang2012}.

Computer simulations indicate that models based on inefficient stellar feedback
(``low''-feedback models) cannot reproduce the level of \ion{O}{vi} column
densities observed in the CGM to impact parameters of 150--200 kpc
\citep[e.g.][]{Stinson2012,Stinson2013,Hummels2013}. The
simulations presented in this work were stopped at \z, but in another project
(Sommer-Larsen \& Laursen 2018, in prep.),
the same code was used to simulate the formation and evolution of eight field
disk galaxies to $z=0$ in a cosmological context using the zoom-in technique.
At $z=0$, the galaxy luminosities, stellar masses, and specific SFRs span the
ranges
$L \sim 0.2\text{--}0.6~L^\star$,
$M_\star \sim 3\text{--}6\e{10}\Msun$, and
$\mathrm{sSFR} \sim 0.04\text{--}0.1\,\mathrm{Gyr}^{-1}$, respectively.

\begin{figure}
\centering
\includegraphics[width=0.48\textwidth]{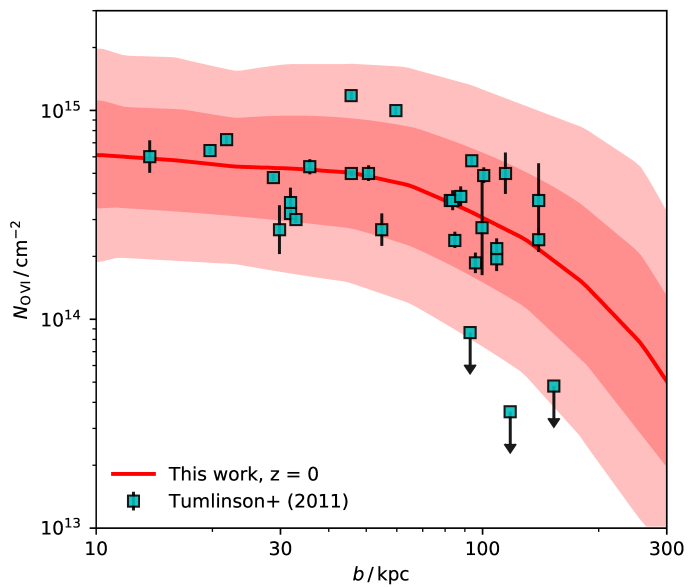}
\caption{Median \ion{O}{vi} column density \NOVI\ (\emph{red}) as a function of
    impact parameter $b$, averaged over LOSs through eight $z=0$ disk galaxies
    simulated using the same code as the one utilized in this work.
    \emph{Shaded} region indicate the 1$\sigma$ and 2$\sigma$ deviations from
    the median, respectively. The observations of ``star-forming'' galaxies by
    \citet{Tumlinson2011} are shown by \emph{cyan squares}.
\label{fig:ovi}}
\end{figure}

For each of the eight galaxies, 6800 LOSs are shot through the galaxy in random
directions and with impact parameters $b$ spanning the range $b=2\text{--}500$
kpc. Along each LOS the total \ion{O}{vi} column density \NOVI\ is determined.
Averaging
over the eight galaxies, \fig{ovi} shows the median \NOVI\ as a function of
impact parameter $b$, compared to the observational data from
\citet{Tumlinson2011} for ``star-forming'' galaxies ($\mathrm{sSFR} \gtrsim
0.01\,\mathrm{Gyr}^{-1}$) of luminosities in the range $L \sim
0.1\text{--}1~L^\star$. As can be seen from the figure, the match between
simulations and observations is quite acceptable, indicating that the
simulations capture at least these outflow properties correctly --- see also
\citet{Sommer-Larsen2017} for the case of the high-impact parameter and
high-metallicity DLA towards Q0918+1636, at $z=2.58$; this system is very well
reproduced by simulations based on a code similar to the one utilized in this
work, and it is shown that by far the majority of the $\alpha$-element
absorption seen in this system at galacto-centric distances
$\gtrsim20\text{--}30$ kpc is caused by metals that have been produced in the
inner part of the galaxy, and subsequently transported outwards by galactic
winds.  

%end outflows

%end disc

\section{Summary and conclusions}
\label{sec:sum}

We have explored numerically the prospects of detecting \lya-emitting galaxies
on the verge of the Epoch of Reionization.
To ensure both good number statistics and high
resolution, we employed a combination of a semi-analytical halo mass funtion
matched to a large-scale $N$-body simulations and zoom-in simulations of
hydrodynamical, cosmological simulations. This was combined with accurate RT
schemes of both ionizing UV and \lya.

Physical uncertainties such as variance in the escape of \lya\ in different
directions --- both out through the ISM of the galaxies and further through the
IGM --- galaxy-to-galaxy variation, and statistical number variance, as well as
observational uncertainties such as differing filter shapes and varying
detector exposure, were taken into account. In general, such uncertainties are
asymmetrical, and a numerical scheme for adding asymmetrical errors is
described (in App.~\ref{app:addAsym}) and made public.

With these simulations we have been able to predict a number of physical
characteristics of galaxies at redshift \z. Our stellar mass--halo mass
relation is consistent with the observationally based model by
\citet{Behroozi2013}, extrapolated to \z. At the investigated redshift the
\lya\ LF is largely unconstrained, but the predicted UV LF is shown to be
consistent with observations. Predicting \lya\ LFs at
lower redshifts where observations are less sparse is beyond the scope of this
work, but will be addressed in future work based on the simulation code
presented here and in \citet{Sommer-Larsen2017}.

Our main findings --- that all apply to \z\ --- are:
\begin{itemize}
\item[$\bullet$]
  Scattering in the CGM prevents \lya\ from escaping a galaxy until large
  distance from its emission site. This results in of the order of half of the
  flux falling outside standard apertures. However, since larger apertures also
  implies more noise, it does not help to enlarge it, and in fact we find that,
  for the \uv, an aperture of 1\farcs4 results in a slightly larger probability
  of detecting a galaxy.
\item[$\bullet$]
  Scattering in the IGM erases completely the blue part of the spectrum, and a
  large fraction ($\sim25\%$) of the red half. This means that, \emph{on
  average}, the \lya\ line is suppressed to continuum level.
\item[$\bullet$]
  The escape of \lya\ is, however, very dependent on the direction from
  which a galaxy is observed. Both the anisotropy of the ISM, and
  inhomogeneities in the transmittance of the IGM, results in the spectra
  exhibiting prominent peaks in a significant fraction of directions, enabling
  us to see them in some cases (see \fig{spec}).
\item [$\bullet$]
  Accurate treatment of the \lya\ RT and the halo number statistics enables of
  to predict the expected number of LAEs in the \uv\ survey; the probabilities
  of detecting 1, 2, or two or more LAEs are
% $\{P(0), P(1), P(2+) \} = \{87_{-47}^{+12},
%                             12_{-11}^{+25},
%                              1_{- 1}^{+16} \}$\%.
  roughly 90\%, 10\%, and 1\%, respectively (see \eq{P_N}).
\item[$\bullet$]
  Since the \lya\ break lies far to the blue in the $J$ broadband filter, and
  since the damping wing of the IGM absorption extends far into the red wing of
  \lya, the IGM suppresses a much smaller fraction in BB than in the NB. Thus,
  a colour selection criterion of $\mathrm{NB} - \mathrm{BB} \lesssim 0$ will
  miss many possible LAEs. Simulating the colours of our galaxies \emph{without}
  the \lya\ emission lines, we show that the lower 1$\sigma$ limit lies
  approximately at $\mathrm{NB} - \mathrm{BB} = +0.85$, which we argue is a
  better criterion.
\item[$\bullet$]
  Using that criterion, we find that even if a galaxy is detectable in the NB,
  its has only $\lesssim35\%$ chance of being selected as a LAE.
\item[$\bullet$]
  We investigated what might lead to a larger probability of success. Using
  a Planck cosmology rather than a WMAP7 increases the predicted number of
  detections from $\sim0.12$ to $\sim0.61$. Assuming an unrealistically less
  aggressive IGM (simply removing the blue wing of the \lya\ line)
  further increase the expected number to 1, while neglecting the IGM
  altogether would result in a handful of detections.
\end{itemize}

% On the whole, while the \uv\ broadband survey has already yielded interesting
% observations \citep{Bowler2012,Bowler2014,Muzzin2013,Ilbert2013}, the hunt
% for LAEs seems unprofitable.

On the whole, while the \uv\ (broadband) survey
has already resulted more than 100 papers directly using the
data\footnote{\href{http://telbib.eso.org/?programid=179.A-2005}
{\tt{http://telbib.eso.org/?programid=179.A-2005}}.}
and many more papers using the data via the
photometric, photo-$z$ and stellar mass catalogues
\citep{Ilbert2013,Muzzin2013,Laigle2016},
the hunt for LAEs seems unprofitable.

According to our simulations, increasing the probable number of detections to
1 (2) would require roughly another 500 (1000) hours of observations with \uv.
However, as the analysis is still in progress, and as our simulations show that
galaxies sufficiently bright for being detected with \uv\ do indeed exist,
hoping is still allowed.

%%%%%%%%%%%%%%%%%%%%%
% Future \lya\ analysis at lower redshifts.....
%
% Lya analysis of lower redshift
% galaxies, based on the simulation code presented here and in SL & Fynbo (2017)
% will be presented in forthcoming papers.
%%%%%%%%%%%%%%%%%%%%%

%end sum

%\section*{Acknowledgements}
%\label{app:ackbib}
\begin{acknowledgements}

The Dark Cosmology Centre was funded by the Danish National Research Foundation
(DNRF).
The Cosmic Dawn Center is funded by DNRF.
PL, BM-J, and JPUF acknowledge support from the ERC-StG grant EGGS-278202. 
Galaxy simulations were performed on the facilities provided by High
Performance Computing Centre at the University of Copenhagen.
Stellar ionization calculations were performed on the {\it kraken} cluster of
the Shared Hierarchical Academic Research Computing Network\footnote{SHARCNET:
\href{http://www.sharcnet.ca}{\tt{www.sharcnet.ca}}.}
In the process of writing our HMF calculator, comparisons with results from
the online HMF calculator
{\sc HMFcalc}\footnote{\href{http://hmf.icrar.org}{\tt{hmf.icrar.org}}.}
\citep{Murray2013} have been extremely helpful.

\end{acknowledgements}
%end ackbib

\bibliographystyle{aa}
\bibliography{./ms_eps}

\begin{thebibliography}{131}
\expandafter\ifx\csname natexlab\endcsname\relax\def\natexlab#1{#1}\fi

\bibitem[{Arnaboldi {et~al.}(2017)Arnaboldi, Gadotti, Hilker, Mascetti, Micol,
  Petr-gotzens, Rejkuba, Ivison, Leibundgut, \& Romaniello}]{Arnaboldi2017}
Arnaboldi, M., Gadotti, D., Hilker, M., {et~al.} 2017, 15

\bibitem[{Barkana \& Loeb(2001)}]{Barkana2001}
Barkana, R. \& Loeb, A. 2001, Phys. Rep., 349, 125

\bibitem[{Barlow(2003)}]{Barlow2003}
Barlow, R. 2003, arXiv:physics/0306138 [\eprint[arXiv]{0306138}]

\bibitem[{Barnes {et~al.}(2011)Barnes, Haehnelt, Tescari, \& Viel}]{Barnes2011}
Barnes, L.~A., Haehnelt, M.~G., Tescari, E., \& Viel, M. 2011, Mon. Not. R.
  Astron. Soc., 416, 1723

\bibitem[{Behroozi {et~al.}(2013)Behroozi, Wechsler, \& Conroy}]{Behroozi2013}
Behroozi, P.~S., Wechsler, R.~H., \& Conroy, C. 2013, Astrophys. J., 770, 57

\bibitem[{Bennett {et~al.}(2013)Bennett, Larson, Weiland, Jarosik, Hinshaw,
  Odegard, Smith, Hill, Gold, Halpern, Komatsu, Nolta, Page, Spergel, Wollack,
  Dunkley, Kogut, Limon, Meyer, Tucker, \& Wright}]{Bennett2013}
Bennett, C.~L., Larson, D., Weiland, J.~L., {et~al.} 2013, Astrophys. J. Suppl.
  Ser., 208, 20

\bibitem[{Binney \& Tremaine(2008)}]{Binney2008}
Binney, J. \& Tremaine, S. 2008, {Galactic Dynamics}, second edi edn., Vol.~1
  (Princeton University Press, Princeton, NJ USA)

\bibitem[{Bond {et~al.}(1991)Bond, Cole, Efstathiou, \& Kaiser}]{Bond1991}
Bond, J.~R., Cole, S., Efstathiou, G., \& Kaiser, N. 1991, Astrophys. J., 379,
  440

\bibitem[{Bouch{\'{e}} {et~al.}(2010)Bouch{\'{e}}, Dekel, Genzel, Genel,
  Cresci, {F{\"{o}}rster Schreiber}, Shapiro, Davies, \& Tacconi}]{Bouche2010}
Bouch{\'{e}}, N., Dekel, A., Genzel, R., {et~al.} 2010, Astrophys. J., 718,
  1001

\bibitem[{Bouwens {et~al.}(2015)Bouwens, Illingworth, Oesch, Trenti,
  Labb{\'{e}}, Bradley, Carollo, van Dokkum, Gonzalez, Holwerda, Franx,
  Spitler, Smit, \& Magee}]{Bouwens2015}
Bouwens, R.~J., Illingworth, G.~D., Oesch, P.~a., {et~al.} 2015, Astrophys. J.,
  803, 34

\bibitem[{Bowman {et~al.}(2018)Bowman, Rogers, Monsalve, Mozdzen, \&
  Mahesh}]{Bowman2018}
Bowman, J.~D., Rogers, A.~E., Monsalve, R.~A., Mozdzen, T.~J., \& Mahesh, N.
  2018, Nature, 555, 67

\bibitem[{Carroll {et~al.}(1992)Carroll, Press, \& Turner}]{Carroll1992}
Carroll, S.~M., Press, W.~H., \& Turner, E.~L. 1992, Annu. Rev. Astron.
  Astrophys., 30, 499

\bibitem[{Chabrier(2003)}]{Chabrier2003}
Chabrier, G. 2003, Publ. Astron. Soc. Pacific, 115, 763

\bibitem[{Cole {et~al.}(2000)Cole, Lacey, Baugh, \& Frenk}]{Cole2000}
Cole, S., Lacey, C.~G., Baugh, C.~M., \& Frenk, C.~S. 2000, Mon. Not. R.
  Astron. Soc., 319, 168

\bibitem[{Crocce {et~al.}(2010)Crocce, Fosalba, Castander, \&
  Gazta{\~{n}}aga}]{Crocce2010}
Crocce, M., Fosalba, P., Castander, F.~J., \& Gazta{\~{n}}aga, E. 2010, Mon.
  Not. R. Astron. Soc., 403, 1353

\bibitem[{Cuby {et~al.}(2007)Cuby, Hibon, Lidman, {Le F{\`{e}}vre}, Gilmozzi,
  Moorwood, \& van~der Werf}]{Cuby2007}
Cuby, J.-G., Hibon, P., Lidman, C., {et~al.} 2007, Astron. Astrophys., 461, 911

\bibitem[{Cui {et~al.}(2012)Cui, Borgani, Dolag, Murante, \&
  Tornatore}]{Cui2012}
Cui, W., Borgani, S., Dolag, K., Murante, G., \& Tornatore, L. 2012, Mon. Not.
  R. Astron. Soc., 423, 2279

\bibitem[{Dijkstra(2014)}]{Dijkstra2014}
Dijkstra, M. 2014, Publ. Astron. Soc. Aust., 31, e040

\bibitem[{Dijkstra {et~al.}(2006)Dijkstra, Haiman, \& Spaans}]{Dijkstra2006}
Dijkstra, M., Haiman, Z.~Z., \& Spaans, M. 2006, Astrophys. J., 649, 14

\bibitem[{Dijkstra \& Loeb(2009)}]{Dijkstra2009}
Dijkstra, M. \& Loeb, A. 2009, Mon. Not. R. Astron. Soc., 400, 1109

\bibitem[{Dijkstra {et~al.}(2011)Dijkstra, Mesinger, \& Wyithe}]{Dijkstra2011}
Dijkstra, M., Mesinger, A., \& Wyithe, J. S.~B. 2011, Mon. Not. R. Astron.
  Soc., 414, 2139

\bibitem[{Dijkstra \& Wyithe(2010)}]{Dijkstra2010}
Dijkstra, M. \& Wyithe, J. S.~B. 2010, Mon. Not. R. Astron. Soc., 408, 352

\bibitem[{Dutton \& {Van Den Bosch}(2009)}]{Dutton2009}
Dutton, A.~A. \& {Van Den Bosch}, F.~C. 2009, Mon. Not. R. Astron. Soc., 396,
  141

\bibitem[{Duval {et~al.}(2014)Duval, Schaerer, {\"{O}}stlin, \&
  Laursen}]{Duval2014}
Duval, F., Schaerer, D., {\"{O}}stlin, G., \& Laursen, P. 2014, Astron.
  Astrophys., 562, A52

\bibitem[{Faisst {et~al.}(2014)Faisst, Capak, Carollo, Scarlata, \&
  Scoville}]{Faisst2014}
Faisst, A.~L., Capak, P., Carollo, C.~M., Scarlata, C., \& Scoville, N. 2014,
  Astrophys. J., 788 [\eprint[arXiv]{1402.3604}]

\bibitem[{Fan {et~al.}(2005)Fan, Strauss, Becker, White, Gunn, Knapp, Richards,
  Schneider, Brinkmann, \& Fukugita}]{Fan2006}
Fan, X., Strauss, M.~A., Becker, R.~H., {et~al.} 2005, Astron. J., 132, 117

\bibitem[{Fardal {et~al.}(2001)Fardal, Katz, Gardner, Hernquist, Weinberg, \&
  Dave}]{Fardal2001}
Fardal, M.~a., Katz, N., Gardner, J.~P., {et~al.} 2001, Astrophys. J., 562, 605

\bibitem[{Faucher-Giguere {et~al.}(2010)Faucher-Giguere, Keres, Dijkstra,
  Hernquist, \& Zaldarriaga}]{Faucher-Giguere2010}
Faucher-Giguere, C.-A., Keres, D., Dijkstra, M., Hernquist, L., \& Zaldarriaga,
  M. 2010, Astrophys. J., 725, 633

\bibitem[{Finkelstein {et~al.}(2013)Finkelstein, Papovich, Dickinson, Song,
  Tilvi, Koekemoer, Finkelstein, Mobasher, Ferguson, Giavalisco, Reddy, Ashby,
  Dekel, Fazio, Fontana, Grogin, Huang, Kocevski, Rafelski, Weiner, \&
  Willner}]{Finkelstein2013}
Finkelstein, S.~L., Papovich, C., Dickinson, M., {et~al.} 2013, Nature, 502,
  524

\bibitem[{Furlanetto {et~al.}(2006)Furlanetto, Zaldarriaga, \&
  Hernquist}]{Furlanetto2006}
Furlanetto, S.~R., Zaldarriaga, M., \& Hernquist, L. 2006, Mon. Not. R. Astron.
  Soc., 365, 1012

\bibitem[{Gehrels(1986)}]{Gehrels1986}
Gehrels, N. 1986, ApJ, 303, 336

\bibitem[{Goerdt \& Ceverino(2015)}]{Goerdt2015}
Goerdt, T. \& Ceverino, D. 2015, Mon. Not. R. Astron. Soc., 450, 3359

\bibitem[{Goerdt {et~al.}(2010)Goerdt, Dekel, Sternberg, Ceverino, Teyssier, \&
  Primack}]{Goerdt2010}
Goerdt, T., Dekel, A., Sternberg, A., {et~al.} 2010, Mon. Not. R. Astron. Soc.,
  407, 613

\bibitem[{Governato {et~al.}(1999)Governato, Babul, Quinn, Tozzi, Baugh, Katz,
  \& Lake}]{Governato1999}
Governato, F., Babul, A., Quinn, T., {et~al.} 1999, Mon. Not. R. Astron. Soc.,
  307, 949

\bibitem[{Gronke {et~al.}(2015)Gronke, Dijkstra, Trenti, \&
  Wyithe}]{Gronke2015a}
Gronke, M., Dijkstra, M., Trenti, M., \& Wyithe, S. 2015, Mon. Not. R. Astron.
  Soc., 449, 1284

\bibitem[{Gupta {et~al.}(2012)Gupta, Mathur, Krongold, Nicastro, \&
  Galeazzi}]{Gupta2012}
Gupta, A., Mathur, S., Krongold, Y., Nicastro, F., \& Galeazzi, M. 2012,
  Astrophys. J. Lett., 756, 4

\bibitem[{Haardt \& Madau(1996)}]{Haardt1996}
Haardt, F. \& Madau, P. 1996, Astrophys. J., 461, 20

\bibitem[{Haardt \& Madau(2012)}]{Haardt2012}
Haardt, F. \& Madau, P. 2012, Astrophys. J., 746 [\eprint[arXiv]{1105.2039}]

\bibitem[{Hansen \& {Peng Oh}(2006)}]{Hansen2006}
Hansen, M. \& {Peng Oh}, S. 2006, Mon. Not. R. Astron. Soc., 367, 979

\bibitem[{Harrington(1973)}]{Harrington1973}
Harrington, J.~P. 1973, Mon. Not. R. Astron. Soc., 162, 43

\bibitem[{Hashimoto {et~al.}(2018)Hashimoto, Laporte, Mawatari, Ellis, Inoue,
  Zackrisson, Roberts-borsani, Zheng, Tamura, Bauer, Fletcher, \&
  Harikane}]{Hashimoto2018}
Hashimoto, T., Laporte, N., Mawatari, K., {et~al.} 2018, Nature, 557, 392

\bibitem[{Hayes {et~al.}(2013)Hayes, {\"{O}}stlin, Schaerer, Verhamme,
  Mas-Hesse, Adamo, Atek, Cannon, Duval, Guaita, Herenz, Kunth, Laursen,
  Melinder, Orlitov{\'{a}}, Ot{\'{i}}-Floranes, \& Sandberg}]{Hayes2013}
Hayes, M., {\"{O}}stlin, G., Schaerer, D., {et~al.} 2013, Astrophys. J., 765,
  L27

\bibitem[{Hayward {et~al.}(2016)Hayward, Hopkins, \& Nov}]{Hayward2016}
Hayward, C.~C., Hopkins, P.~F., \& Nov, G.~A. 2016, 16, 1

\bibitem[{Heckman {et~al.}(2015)Heckman, Alexandroff, Borthakur, Overzier, \&
  Leitherer}]{Heckman2015}
Heckman, T.~M., Alexandroff, R.~M., Borthakur, S., Overzier, R., \& Leitherer,
  C. 2015, Astrophys. J., 809, 147

\bibitem[{Hu {et~al.}(2016)Hu, Cowie, Songaila, Barger, Rosenwasser, \&
  Wold}]{Hu2016}
Hu, E.~M., Cowie, L.~L., Songaila, A., {et~al.} 2016, 7, 5

\bibitem[{Hummels {et~al.}(2013)Hummels, Bryan, Smith, \& Turk}]{Hummels2013}
Hummels, C.~B., Bryan, G.~L., Smith, B.~D., \& Turk, M.~J. 2013, Mon. Not. R.
  Astron. Soc., 430, 1548

\bibitem[{Ilbert {et~al.}(2013)Ilbert, McCracken, Fevre, Capak, Dunlop, Karim,
  Renzini, Caputi, Boissier, Arnouts, Aussel, Comparat, Guo, Hudelot,
  Kartaltepe, Kneib, Krogager, Floc'h, Lilly, Mellier, Milvang-Jensen, Moutard,
  Onodera, Richard, Salvato, Sanders, Scoville, Silverman, Taniguchi, Tasca,
  Thomas, Toft, Tresse, Vergani, Wolk, \& Zirm}]{Ilbert2013}
Ilbert, O., McCracken, H.~J., Fevre, O.~L., {et~al.} 2013, Astron. Astrophys.,
  556, A55

\bibitem[{Iliev {et~al.}(2008)Iliev, Shapiro, McDonald, Mellema, \&
  Pen}]{Iliev2008}
Iliev, I.~T., Shapiro, P.~R., McDonald, P., Mellema, G., \& Pen, U.-L. 2008,
  Mon. Not. R. Astron. Soc., 391, 63

\bibitem[{Iye {et~al.}(2006)Iye, Ota, Kashikawa, Furusawa, Hashimoto, Hattori,
  Matsuda, Morokuma, Ouchi, \& Shimasaku}]{Iye2006}
Iye, M., Ota, K., Kashikawa, N., {et~al.} 2006, Nature, 443, 186

\bibitem[{Jarosik {et~al.}(2011)Jarosik, Bennett, Dunkley, Gold, Greason,
  Halpern, Hill, Hinshaw, Kogut, Komatsu, Larson, Limon, Meyer, Nolta, Odegard,
  Page, Smith, Spergel, Tucker, Weiland, Wollack, \& Wright}]{Jarosik2010}
Jarosik, N., Bennett, C.~L., Dunkley, J., {et~al.} 2011, Astrophys. Journal,
  Suppl. Ser., 192, 14

\bibitem[{Jensen {et~al.}(2013)Jensen, Laursen, Mellema, Iliev, Sommer-Larsen,
  \& Shapiro}]{Jensen2013}
Jensen, H., Laursen, P., Mellema, G., {et~al.} 2013, Mon. Not. R. Astron. Soc.,
  428, 1366

\bibitem[{Kennicutt(1998)}]{Kennicutt1998}
Kennicutt, R.~C. 1998, Annu. Rev. Astron. Astrophys., 36, 189

\bibitem[{Klypin {et~al.}(2016)Klypin, Yepes, Gottl{\"{o}}ber, Prada, \&
  He{\ss}}]{Klypin2016}
Klypin, A., Yepes, G., Gottl{\"{o}}ber, S., Prada, F., \& He{\ss}, S. 2016,
  Mon. Not. R. Astron. Soc., 457, 4340

\bibitem[{Klypin {et~al.}(2011)Klypin, Trujillo-Gomez, \& Primack}]{Klypin2011}
Klypin, A.~A., Trujillo-Gomez, S., \& Primack, J. 2011, Astrophys. J., 740, 102

\bibitem[{Kunth {et~al.}(1998)Kunth, Mas-Hesse, Terlevich, Terlevich, Lequeux,
  \& Fall}]{Kunth1998}
Kunth, D., Mas-Hesse, J.~M., Terlevich, E., {et~al.} 1998, Astron. Astrophys.,
  334, 11

\bibitem[{Lahav {et~al.}(1991)Lahav, Lilje, Primack, \& Rees}]{Lahav1991}
Lahav, O., Lilje, P.~B., Primack, J.~R., \& Rees, M.~J. 1991, Mon. Not. R.
  Astron. Soc., 251, 128

\bibitem[{Laigle {et~al.}(2016)Laigle, McCracken, Ilbert, Hsieh, Davidzon,
  Capak, Hasinger, Silverman, Pichon, Coupon, Aussel, Borgne, Caputi, Cassata,
  Chang, Civano, Dunlop, Fynbo, Kartaltepe, Koekemoer, Fevre, Floc'h,
  Leauthaud, Lilly, Lin, Marchesi, Milvang-Jensen, Salvato, Sanders, Scoville,
  Smolcic, Stockmann, Taniguchi, Tasca, Toft, Vaccari, \& Zabl}]{Laigle2016}
Laigle, C., McCracken, H.~J., Ilbert, O., {et~al.} 2016, Astrophys. J. Suppl.
  Ser., 224, 1

\bibitem[{Laursen {et~al.}(2013)Laursen, Duval, \& {\"{O}}stlin}]{Laursen2013}
Laursen, P., Duval, F., \& {\"{O}}stlin, G. 2013, Astrophys. J., 766, 124

\bibitem[{Laursen {et~al.}(2009{\natexlab{a}})Laursen, Razoumov, \&
  Sommer-Larsen}]{Laursen2009a}
Laursen, P., Razoumov, A.~O., \& Sommer-Larsen, J. 2009{\natexlab{a}},
  Astrophys. J., 696, 853

\bibitem[{Laursen {et~al.}(2009{\natexlab{b}})Laursen, Sommer-Larsen, \&
  Andersen}]{Laursen2009b}
Laursen, P., Sommer-Larsen, J., \& Andersen, A.~C. 2009{\natexlab{b}},
  Astrophys. J., 704, 1640

\bibitem[{Laursen {et~al.}(2011)Laursen, Sommer-Larsen, \&
  Razoumov}]{Laursen2011}
Laursen, P., Sommer-Larsen, J., \& Razoumov, A.~O. 2011, Astrophys. J., 728, 52

\bibitem[{{Le Delliou} {et~al.}(2006){Le Delliou}, Lacey, Baugh, \&
  Morris}]{Delliou2006}
{Le Delliou}, M., Lacey, C.~G., Baugh, C.~M., \& Morris, S.~L. 2006, Mon. Not.
  R. Astron. Soc., 365, 712

\bibitem[{Leclercq {et~al.}(2017)Leclercq, Bacon, Wisotzki, Mitchell, Garel,
  Verhamme, Blaizot, Hashimoto, Herenz, Conseil, Cantalupo, Inami, Contini,
  Richard, Maseda, Schaye, Marino, Akhlaghi, Brinchmann, \&
  Carollo}]{Leclercq2017}
Leclercq, F., Bacon, R., Wisotzki, L., {et~al.} 2017
  [\eprint[arXiv]{1710.10271}]

\bibitem[{Leitherer {et~al.}(1999)Leitherer, Schaerer, Goldader, Gonza,
  Delgado, {Foo Kune}, {De Mello}, Devost, \& Heckman}]{Leitherer1999}
Leitherer, C., Schaerer, D., Goldader, J.~D., {et~al.} 1999, Astrophys. J.
  Suppl. Ser., 123, 3

\bibitem[{Martizzi {et~al.}(2014)Martizzi, Mohammed, Teyssier, \&
  Moore}]{Martizzi2014}
Martizzi, D., Mohammed, I., Teyssier, R., \& Moore, B. 2014, Mon. Not. R.
  Astron. Soc., 440, 2290

\bibitem[{Matthee {et~al.}(2018)Matthee, Sobral, Gronke, Paulino-Afonso,
  Stefanon, \& R{\"{o}}ttgering}]{Matthee2018}
Matthee, J., Sobral, D., Gronke, M., {et~al.} 2018, 1

\bibitem[{Matthee {et~al.}(2014)Matthee, Sobral, Swinbank, Smail, Best, Kim,
  Franx, Milvang-Jensen, \& Fynbo}]{Matthee2014}
Matthee, J.~J., Sobral, D., Swinbank, A.~M., {et~al.} 2014, Mon. Not. R.
  Astron. Soc., 440, 2375

\bibitem[{McCracken {et~al.}(2012)McCracken, Milvang-Jensen, Dunlop, Franx,
  Fynbo, {Le F{\`{e}}vre}, Holt, Caputi, Goranova, Buitrago, Emerson,
  Freudling, Hudelot, L{\'{o}}pez-Sanjuan, Magnard, Mellier, M{\o}ller,
  Nilsson, Sutherland, Tasca, \& Zabl}]{McCracken2012}
McCracken, H.~J., Milvang-Jensen, B., Dunlop, J., {et~al.} 2012, Astron.
  Astrophys., 544, A156

\bibitem[{McQuinn {et~al.}(2007)McQuinn, Hernquist, Zaldarriaga, \&
  Dutta}]{McQuinn2007}
McQuinn, M., Hernquist, L., Zaldarriaga, M., \& Dutta, S. 2007, Mon. Not. R.
  Astron. Soc., 381, 75

\bibitem[{Milvang-Jensen {et~al.}(2013)Milvang-Jensen, Freudling, Zabl, Fynbo,
  M{\o}ller, Nilsson, McCracken, Hjorth, {Le F{\`{e}}vre}, Tasca, Dunlop, \&
  Sobral}]{Milvang-Jensen2013}
Milvang-Jensen, B., Freudling, W., Zabl, J., {et~al.} 2013, Astron. Astrophys.,
  560, A94

\bibitem[{Muldrew {et~al.}(2017)Muldrew, Hatch, \& Cooke}]{Muldrew2017}
Muldrew, S.~I., Hatch, N.~A., \& Cooke, E.~A. 2017, 14, 1

\bibitem[{Muratov {et~al.}(2015)Muratov, Kere{\v{s}}, Faucher-Gigu{\`{e}}re,
  Hopkins, Quataert, \& Murray}]{Muratov2015}
Muratov, A.~L., Kere{\v{s}}, D., Faucher-Gigu{\`{e}}re, C.~A., {et~al.} 2015,
  Mon. Not. R. Astron. Soc., 454, 2691

\bibitem[{Murray {et~al.}(2013)Murray, Power, \& Robotham}]{Murray2013}
Murray, S., Power, C., \& Robotham, A. 2013, Astron. Comput., 3-4, 23

\bibitem[{Muzzin {et~al.}(2013)Muzzin, Marchesini, Stefanon, Franx,
  Milvang-Jensen, Dunlop, Fynbo, Brammer, Labb{\'{e}}, \& van
  Dokkum}]{Muzzin2013}
Muzzin, A., Marchesini, D., Stefanon, M., {et~al.} 2013, Astrophys. J. Suppl.
  Ser., 206, 8

\bibitem[{Navarro \& White(1994)}]{Navarro1994}
Navarro, J.~F. \& White, S. D.~M. 1994, Mon. Not. R. Astron. Soc., 267, 401

\bibitem[{Neufeld(1990)}]{Neufeld1990}
Neufeld, D.~A. 1990, Astrophys. J., 350, 216

\bibitem[{Neufeld(1991)}]{Neufeld1991}
Neufeld, D.~A. 1991, Astrophys. J., 370, L85

\bibitem[{Nilsson {et~al.}(2007)Nilsson, Orsi, Lacey, Baugh, \&
  Thommes}]{Nilsson2007}
Nilsson, K.~K., Orsi, A., Lacey, C.~G., Baugh, C.~M., \& Thommes, E. 2007,
  Astron. Astrophys., 474, 385

\bibitem[{Oesch {et~al.}(2017)Oesch, Bouwens, Illingworth, Labbe, \&
  Stefanon}]{Oesch2017}
Oesch, P.~A., Bouwens, R.~J., Illingworth, G.~D., Labbe, I., \& Stefanon, M.
  2017 [\eprint[arXiv]{1710.11131}]

\bibitem[{Oesch {et~al.}(2016)Oesch, Brammer, van Dokkum, Illingworth, Bouwens,
  Labb{\'{e}}, Franx, Momcheva, Ashby, Fazio, Gonzalez, Holden, Magee, Skelton,
  Smit, Spitler, Trenti, \& Willner}]{Oesch2016}
Oesch, P.~A., Brammer, G., van Dokkum, P.~G., {et~al.} 2016, Astrophys. J.,
  819, 129

\bibitem[{O{\~{n}}orbe {et~al.}(2013)O{\~{n}}orbe, Garrison-Kimmel, Maller,
  Bullock, Rocha, \& Hahn}]{Onorbe2013}
O{\~{n}}orbe, J., Garrison-Kimmel, S., Maller, A.~H., {et~al.} 2013, Mon. Not.
  R. Astron. Soc., 437, 1894

\bibitem[{Partridge \& Peebles(1967)}]{Partridge1967}
Partridge, R.~B. \& Peebles, P. J.~E. 1967, Astrophys. J., 147, 868

\bibitem[{Pettini {et~al.}(2001)Pettini, Shapley, Steidel, Cuby, Dickinson,
  Moorwood, Adelberger, \& Giavalisco}]{Pettini2001}
Pettini, M., Shapley, A.~E., Steidel, C.~C., {et~al.} 2001, Astrophys. J., 554,
  981

\bibitem[{{Planck Collaboration:} {et~al.}(2016){Planck Collaboration:}, Ade,
  Aghanim, Arnaud, Ashdown, Aumont, Baccigalupi, Banday, Barreiro, Bartlett,
  Bartolo, Battaner, Battye, Benabed, Beno{\^{i}}t, Benoit-L{\'{e}}vy, Bernard,
  Bersanelli, Bielewicz, Bock, Bonaldi, Bonavera, Bond, Borrill, Bouchet,
  Boulanger, Bucher, Burigana, Butler, Calabrese, Cardoso, Catalano, Challinor,
  Chamballu, Chary, Chiang, Chluba, Christensen, Church, Clements, Colombi,
  Colombo, Combet, Coulais, Crill, Curto, Cuttaia, Danese, Davies, Davis,
  de~Bernardis, de~Rosa, de~Zotti, Delabrouille, D{\'{e}}sert, {Di Valentino},
  Dickinson, Diego, Dolag, Dole, Donzelli, Dor{\'{e}}, Douspis, Ducout,
  Dunkley, Dupac, Efstathiou, Elsner, En{\ss}lin, Eriksen, Farhang, Fergusson,
  Finelli, Forni, Frailis, Fraisse, Franceschi, Frejsel, Galeotta, Galli,
  Ganga, Gauthier, Gerbino, Ghosh, Giard, Giraud-H{\'{e}}raud, Giusarma,
  Gjerl{\o}w, Gonz{\'{a}}lez-Nuevo, G{\'{o}}rski, Gratton, Gregorio, Gruppuso,
  Gudmundsson, Hamann, Hansen, Hanson, Harrison, Helou, Henrot-Versill{\'{e}},
  Hern{\'{a}}ndez-Monteagudo, Herranz, Hildebrandt, Hivon, Hobson, Holmes,
  Hornstrup, Hovest, Huang, Huffenberger, Hurier, Jaffe, Jaffe, Jones, Juvela,
  Keih{\"{a}}nen, Keskitalo, Kisner, Kneissl, Knoche, Knox, Kunz, Kurki-Suonio,
  Lagache, L{\"{a}}hteenm{\"{a}}ki, Lamarre, Lasenby, Lattanzi, Lawrence,
  Leahy, Leonardi, Lesgourgues, Levrier, Lewis, Liguori, Lilje,
  Linden-V{\o}rnle, L{\'{o}}pez-Caniego, Lubin, Mac{\'{i}}as-P{\'{e}}rez,
  Maggio, Maino, Mandolesi, Mangilli, Marchini, Maris, Martin, Martinelli,
  Mart{\'{i}}nez-Gonz{\'{a}}lez, Masi, Matarrese, McGehee, Meinhold,
  Melchiorri, Melin, Mendes, Mennella, Migliaccio, Millea, Mitra,
  Miville-Desch{\^{e}}nes, Moneti, Montier, Morgante, Mortlock, Moss, Munshi,
  Murphy, Naselsky, Nati, Natoli, Netterfield, N{\o}rgaard-Nielsen, Noviello,
  Novikov, Novikov, Oxborrow, Paci, Pagano, Pajot, Paladini, Paoletti,
  Partridge, Pasian, Patanchon, Pearson, Perdereau, Perotto, Perrotta,
  Pettorino, Piacentini, Piat, Pierpaoli, Pietrobon, Plaszczynski,
  Pointecouteau, Polenta, Popa, Pratt, Pr{\'{e}}zeau, Prunet, Puget, Rachen,
  Reach, Rebolo, Reinecke, Remazeilles, Renault, Renzi, Ristorcelli, Rocha,
  Rosset, Rossetti, Roudier, Rouill{\'{e}}~d'Orfeuil, Rowan-Robinson,
  Rubi{\~{n}}o-Mart{\'{i}}n, Rusholme, Said, Salvatelli, Salvati, Sandri,
  Santos, Savelainen, Savini, Scott, Seiffert, Serra, Shellard, Spencer,
  Spinelli, Stolyarov, Stompor, Sudiwala, Sunyaev, Sutton, Suur-Uski, Sygnet,
  Tauber, Terenzi, Toffolatti, Tomasi, Tristram, Trombetti, Tucci, Tuovinen,
  T{\"{u}}rler, Umana, Valenziano, Valiviita, {Van Tent}, Vielva, Villa, Wade,
  Wandelt, Wehus, White, White, Wilkinson, Yvon, Zacchei, \&
  Zonca}]{PlanckCollaboration2016}
{Planck Collaboration:}, Ade, P. A.~R., Aghanim, N., {et~al.} 2016, Astron.
  Astrophys., 594, A13

\bibitem[{Powell {et~al.}(2011)Powell, Slyz, \& Devriendt}]{Powell2011}
Powell, L.~C., Slyz, A., \& Devriendt, J. 2011, Mon. Not. R. Astron. Soc., 414,
  3671

\bibitem[{Prescott {et~al.}(2015)Prescott, Momcheva, Brammer, Fynbo, \&
  M{\o}ller}]{Prescott2015}
Prescott, M. K.~M., Momcheva, I., Brammer, G.~B., Fynbo, J. P.~U., \&
  M{\o}ller, P. 2015, Astrophys. J., 802, 32

\bibitem[{Press \& Schechter(1974)}]{Press1974}
Press, W.~H. \& Schechter, P. 1974, Astrophys. J., 187, 425

\bibitem[{Prochaska {et~al.}(2011)Prochaska, Weiner, Chen, Mulchaey, \&
  Cooksey}]{Prochaska2011}
Prochaska, J.~X., Weiner, B., Chen, H.~W., Mulchaey, J., \& Cooksey, K. 2011,
  Astrophys. J., 740, 1

\bibitem[{Prochaska {et~al.}(2017)Prochaska, Werk, Worseck, Tripp, Tumlinson,
  Burchett, Fox, Fumagalli, Lehner, Peeples, \& Tejos}]{Prochaska2017}
Prochaska, J.~X., Werk, J.~K., Worseck, G., {et~al.} 2017, Astrophys. J., 837,
  169

\bibitem[{Rahmati {et~al.}(2013)Rahmati, Pawlik, Rai{\v{c}}evic, \&
  Schaye}]{Rahmati2013}
Rahmati, A., Pawlik, A.~H., Rai{\v{c}}evic, M., \& Schaye, J. 2013, Mon. Not.
  R. Astron. Soc., 430, 2427

\bibitem[{Raiter {et~al.}(2010)Raiter, Schaerer, \& Fosbury}]{Raiter2010}
Raiter, A., Schaerer, D., \& Fosbury, R. A.~E. 2010, Astron. Astrophys., 523,
  A64

\bibitem[{Razoumov \& Sommer-Larsen(2006)}]{Razoumov2006}
Razoumov, A.~O. \& Sommer-Larsen, J. 2006, Astrophys. J., 651, L89

\bibitem[{Razoumov \& Sommer-Larsen(2007)}]{Razoumov2007}
Razoumov, A.~O. \& Sommer-Larsen, J. 2007, Astrophys. J., 668, 674

\bibitem[{Rodr{\'{i}}guez-Puebla {et~al.}(2016)Rodr{\'{i}}guez-Puebla,
  Behroozi, Primack, Klypin, Lee, \& Hellinger}]{Rodriguez-Puebla2016}
Rodr{\'{i}}guez-Puebla, A., Behroozi, P., Primack, J., {et~al.} 2016, Mon. Not.
  R. Astron. Soc., 462, 893

\bibitem[{Romeo {et~al.}(2006)Romeo, Sommer-Larsen, Portinari, \&
  Antonuccio-Delogu}]{Romeo2006}
Romeo, A.~D., Sommer-Larsen, J., Portinari, L., \& Antonuccio-Delogu, V. 2006,
  Mon. Not. R. Astron. Soc., 371, 548

\bibitem[{Sadoun {et~al.}(2016)Sadoun, Zheng, \&
  Miralda-Escud{\'{e}}}]{Sadoun2016}
Sadoun, R., Zheng, Z., \& Miralda-Escud{\'{e}}, J. 2016, 7, 1

\bibitem[{Salpeter(1955)}]{Salpeter1955}
Salpeter, E.~E. 1955, Astrophys. J., 121, 161

\bibitem[{Schechter(1976)}]{Schechter1976}
Schechter, P. 1976, Astrophys. J., 203, 297

\bibitem[{Schroetter {et~al.}(2015)Schroetter, Bouch{\'{e}}, P{\'{e}}roux,
  Murphy, Contini, \& Finley}]{Schroetter2015}
Schroetter, I., Bouch{\'{e}}, N., P{\'{e}}roux, C., {et~al.} 2015, Astrophys.
  J., 804, 1

\bibitem[{Scoville {et~al.}(2007)Scoville, Aussel, Brusa, Capak, Carollo,
  Elvis, Giavalisco, Guzzo, Hasinger, Impey, Kneib, LeFevre, Lilly, Mobasher,
  Renzini, Rich, Sanders, Schinnerer, Schminovich, Shopbell, Taniguchi, \&
  Tyson}]{Scoville2007}
Scoville, N., Aussel, H., Brusa, M., {et~al.} 2007, Astrophys. J. Suppl. Ser.,
  172, 1

\bibitem[{Shapley {et~al.}(2003)Shapley, Steidel, Pettini, \&
  Adelberger}]{Shapley2003}
Shapley, A.~E., Steidel, C.~C., Pettini, M., \& Adelberger, K.~L. 2003,
  Astrophys. J., 588, 65

\bibitem[{Sheth {et~al.}(2001)Sheth, Mo, \& Tormen}]{Sheth2001}
Sheth, R.~K., Mo, H.~J., \& Tormen, G. 2001, Mon. Not. R. Astron. Soc., 323, 1

\bibitem[{Sheth \& Tormen(1999)}]{Sheth1999}
Sheth, R.~K. \& Tormen, G. 1999, Mon. Not. R. Astron. Soc., 308, 119

\bibitem[{Sheth \& Tormen(2002)}]{Sheth2002}
Sheth, R.~K. \& Tormen, G. 2002, Mon. Not. R. Astron. Soc., 329, 61

\bibitem[{Shibuya {et~al.}(2012)Shibuya, Kashikawa, Ota, Iye, Ouchi, Furusawa,
  Shimasaku, \& Hattori}]{Shibuya2012}
Shibuya, T., Kashikawa, N., Ota, K., {et~al.} 2012, Astrophys. J., 752, 114

\bibitem[{Sobral {et~al.}(2009)Sobral, Best, Geach, Smail, Kurk, Cirasuolo,
  Casali, Ivison, Coppin, \& Dalton}]{Sobral2009}
Sobral, D., Best, P.~N., Geach, J.~E., {et~al.} 2009, Mon. Not. R. Astron. Soc.
  Lett., 398, 68

\bibitem[{Sommer-Larsen \& Fynbo(2017)}]{Sommer-Larsen2017}
Sommer-Larsen, J. \& Fynbo, J. P.~U. 2017, Mon. Not. R. Astron. Soc., 464, 2441

\bibitem[{Sommer-Larsen {et~al.}(2005)Sommer-Larsen, Romeo, \&
  Portinari}]{Sommer-Larsen2005}
Sommer-Larsen, J., Romeo, A.~D., \& Portinari, L. 2005, Mon. Not. R. Astron.
  Soc., 357, 478

\bibitem[{Stanek {et~al.}(2009)Stanek, Rudd, \& Evrard}]{Stanek2009}
Stanek, R., Rudd, D., \& Evrard, A.~E. 2009, Mon. Not. R. Astron. Soc. Lett.,
  394, 11

\bibitem[{Stinson {et~al.}(2013)Stinson, Brook, Macci{\`{o}}, Wadsley, Quinn,
  \& Couchman}]{Stinson2013}
Stinson, G.~S., Brook, C., Macci{\`{o}}, A.~V., {et~al.} 2013, Mon. Not. R.
  Astron. Soc., 428, 129

\bibitem[{Stinson {et~al.}(2012)Stinson, Brook, Prochaska, Hennawi, Shen,
  Wadsley, Pontzen, Couchman, Quinn, Macci??, \& Gibson}]{Stinson2012}
Stinson, G.~S., Brook, C., Prochaska, J.~X., {et~al.} 2012, Mon. Not. R.
  Astron. Soc., 425, 1270

\bibitem[{Sutherland {et~al.}(2014)Sutherland, Emerson, Dalton, Atad-Ettedgui,
  Beard, Bennett, Bezawada, Born, Caldwell, Clark, Craig, Henry, Jeffers,
  Little, McPherson, Murray, Stewart, Stobie, Terrett, Ward, Whalley, \&
  Woodhouse}]{Sutherland2015}
Sutherland, W., Emerson, J., Dalton, G., {et~al.} 2014, Astron. Astrophys.,
  575, A25

\bibitem[{Tapken {et~al.}(2007)Tapken, Appenzeller, Noll, Richling, Heidt,
  Meink{\"{o}}hn, \& Mehlert}]{Tapken2007}
Tapken, C., Appenzeller, I., Noll, S., {et~al.} 2007, 72, 63

\bibitem[{Thiele(1889)}]{Thiele1889}
Thiele, T.~N. 1889, {Forel{\ae}sninger over almindelig Iagttagelsesl{\ae}re:
  Sandsynlighedsregning og mindste Kvadraters Methode} (C. A. Reitzel)

\bibitem[{Thommes \& Meisenheimer(2005)}]{Thommes2005}
Thommes, E. \& Meisenheimer, K. 2005, Astronomy, 891, 877

\bibitem[{Tinker {et~al.}(2008)Tinker, Kravtsov, Klypin, Abazajian, Warren,
  Yepes, Gottlober, \& Holz}]{Tinker2008}
Tinker, J.~L., Kravtsov, A.~V., Klypin, A., {et~al.} 2008, Astrophys. J., 688,
  709

\bibitem[{Trac {et~al.}(2015)Trac, Cen, \& Mansfield}]{Trac2015}
Trac, H., Cen, R., \& Mansfield, P. 2015, Astrophys. J., 813, 54

\bibitem[{Trenti \& Stiavelli(2008)}]{Trenti2008}
Trenti, M. \& Stiavelli, M. 2008

\bibitem[{Tumlinson {et~al.}(2011)Tumlinson, Thom, Werk, Prochaska, Tripp,
  Weinberg, Peeples, O'Meara, Oppenheimer, Meiring, Katz, Dav{\'{e}}, Ford, \&
  Sembach}]{Tumlinson2011}
Tumlinson, J., Thom, C., Werk, J.~K., {et~al.} 2011, Science (80-. )., 334, 948

\bibitem[{Vanzella {et~al.}(2011)Vanzella, Pentericci, Fontana, Grazian,
  Castellano, Boutsia, Cristiani, Dickinson, Gallozzi, Giallongo, Giavalisco,
  Maiolino, Moorwood, Paris, \& Santini}]{Vanzella2011}
Vanzella, E., Pentericci, L., Fontana, A., {et~al.} 2011, Astrophys. J., 730,
  L35

\bibitem[{Verhamme {et~al.}(2012)Verhamme, Dubois, Blaizot, Garel, Bacon,
  Devriendt, Guiderdoni, \& Slyz}]{Verhamme2012}
Verhamme, A., Dubois, Y., Blaizot, J., {et~al.} 2012, Astron. Astrophys., 546,
  A111

\bibitem[{Verhamme {et~al.}(2006)Verhamme, Schaerer, \& Maselli}]{Verhamme2006}
Verhamme, A., Schaerer, D., \& Maselli, A. 2006, Astron. Astrophys., 460, 397

\bibitem[{Vincenzo {et~al.}(2016)Vincenzo, Matteucci, Belfiore, \&
  Maiolino}]{Vincenzo2016}
Vincenzo, F., Matteucci, F., Belfiore, F., \& Maiolino, R. 2016, Mon. Not. R.
  Astron. Soc., 455, 4183

\bibitem[{Visbal {et~al.}(2012)Visbal, Barkana, Fialkov, Tseliakhovich, \&
  Hirata}]{Visbal2012}
Visbal, E., Barkana, R., Fialkov, A., Tseliakhovich, D., \& Hirata, C.~M. 2012,
  Nature, 487, 70

\bibitem[{Wang \& Yao(2012)}]{Wang2012}
Wang, Q.~D. \& Yao, Y. 2012, ArXiv e-prints, 000, 6

\bibitem[{Watson {et~al.}(2013)Watson, Iliev, D'Aloisio, Knebe, Shapiro, \&
  Yepes}]{Watson2013}
Watson, W.~A., Iliev, I.~T., D'Aloisio, A., {et~al.} 2013, Mon. Not. R. Astron.
  Soc., 433, 1230

\bibitem[{Willis \& Courbin(2005)}]{Willis2005}
Willis, J.~P. \& Courbin, F. 2005, Mon. Not. R. Astron. Soc., 357, 1348

\bibitem[{Willis {et~al.}(2008)Willis, Courbin, Kneib, \& Minniti}]{Willis2008}
Willis, J.~P., Courbin, F., Kneib, J.~P., \& Minniti, D. 2008, Mon. Not. R.
  Astron. Soc., 384, 1039

\bibitem[{Yajima {et~al.}(2012)Yajima, Li, Zhu, \& Abel}]{Yajima2012}
Yajima, H., Li, Y., Zhu, Q., \& Abel, T. 2012, Mon. Not. R. Astron. Soc., 424,
  884

\bibitem[{Yamada {et~al.}(2012)Yamada, Matsuda, Kousai, Hayashino, Morimoto, \&
  Umemura}]{Yamada2012}
Yamada, T., Matsuda, Y., Kousai, K., {et~al.} 2012, Astrophys. J., 751, 29

\bibitem[{Zitrin {et~al.}(2015)Zitrin, Labb{\'{e}}, Belli, Bouwens, Ellis,
  Roberts-Borsani, Stark, Oesch, \& Smit}]{Zitrin2015}
Zitrin, A., Labb{\'{e}}, I., Belli, S., {et~al.} 2015, Astrophys. J., 810, L12

\end{thebibliography}
% Turn cccnn key in ref  into AuthorYear, standing at beginning of line: f{di{WywBpF(yi(f}PWi			
% Turn cccnn key in cite into AuthorYear, standing at beginning of ref:  mm*GNWywF(pyi(u`mP de/\<\w\w\w\d\d\>

\begin{appendix}

\section{Convergence studies}
\label{app:conv}

% To have confidence in the obtained results, we must make sure that they do not
% change notably for reasonably small changes in both uncertain physical input
% parameters and numerical factors. \moca\ has already been investigated for some
% of these effects, such as resolution of the underlying hydrosimulation and AMR
% grid, effects of the UV RT schemes, and dust parameters
% \citep{Laursen2009a,Laursen2009b}, as well as reionization kick-in
% $z_{\mathrm{re}}$ and density fluctuation scale $\sigma_8$ \citep{Laursen2011}.
% In the following sections we perform some additional convergence tests.

% \subsection{Numerical resolution}
% \label{sec:numres}
As described in \sec{sims}, for convergence study purposes, seven
simulations were carried out at eight times higher mass resolution and twice
better linear resolution than the production simulations described and utilized
in this work. Unfortunately, the HR regions for these simulations
turned out to be too small to robustly carry out \lya~RT, which by nature is
non-local. Nevertheless, other properties of the galaxies formed can be
compared to the results of the production runs.

\begin{figure}
\centering
\includegraphics[width=0.40\textwidth]{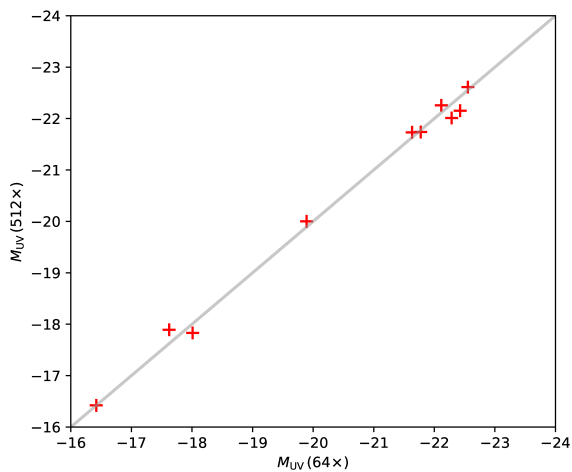}
\caption{Absolute UV magnitudes at 1600 {\AA} $M_\mathrm{UV}$ for the very high-res simulations
    (512$\times$) vs. production runs (64$\times$). No dust-correction has been
    performed.
    To guide the eye, the \emph{grey} line shows a unit slope.
\label{fig:muv512x64x}}
\end{figure}

\Fig{muv512x64x} shows the absolute, dust-uncorrected 1600 {\AA}~AB magnitudes
$M_\mathrm{UV}$ for ten galaxies at 512$\times$ vs. 64$\times$ resolution. As
can be seen from the figure, the correspondence between the 512$\times$ and
64$\times$ magnitudes is quite good, with differences being $\la$ 0.2 mag.

\begin{figure}
\centering
\includegraphics[width=0.40\textwidth]{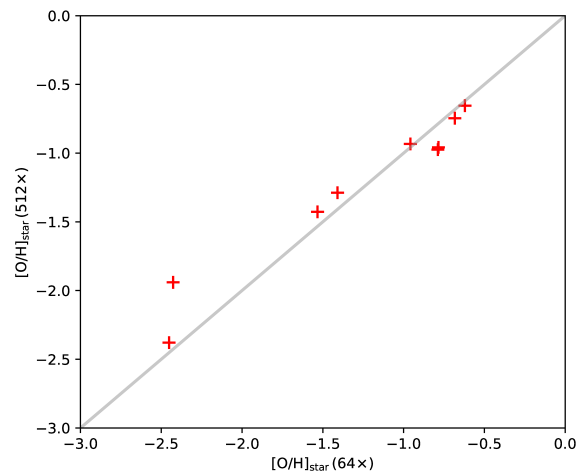}
\caption{Median stellar metallicity [O/H] for nine galaxies: very
    high-resolution simulations (512$\times$) vs. production runs (64$\times$).
    To guide the eye, the \emph{grey} line shows a unit slope.
\label{fig:oh512x64x}}
\end{figure}

\Fig{oh512x64x} shows the median stellar metallicity [O/H] for nine of the
galaxies at 512$\times$ vs. 64$\times$ resolution. For eight of the galaxies,
there is reasonable agreement between 512$\times$ and 64$\times$ values, with
differences being $\lesssim0.2$ dex. For the ninth galaxy, the difference is
somewhat larger, $\sim$0.5 dex. This galaxy is small, however, with the
64$\times$ version consisting of 69 star particles, so stochastic effects are
significant.

These results indicate that convergence has been successfully achieved.

% 
% \subsection{Resolution}
% \label{app:resol}
% 
% 
% 
% %end resol
% 
% \subsection{Cosmology}
% \label{app:cosmo}
% 
% 
% 
% %end cosmo
% 
% \subsection{Outflow models}
% \label{app:models}
% 
% 
% 
% %end models
% 
% \subsection{Threshold between galactic and intergalactic RT}
% \label{app:r0}
% 
% 
% 
% %end r0
% 
% %end conv

\section{Adding asymmetric uncertainties}
\label{app:addAsym}

On several occasions during this work, it was necessary to add numbers with
uncertainties, or errors, not governed by Gaussian probability density
functions (PDFs). That is, numbers that
are written not as $\X = x_0\pm\sigma$, but as
$\X = {x_0}_{-\siglos}^{+\sighis}$, where $x_0$ is the central value, and
\siglo\ and \sighi\ are the lower and upper error, respectively. If the full
PDFs are known, the result $\X + \Y$ can simply be
obtained by convolution or Monte Carlo sampling. In many cases, however, they
are not.  When only the set $\{x_0,\siglo,\sighi\}$ is known, there cannot be a
unique solution, since infinitely many PDFs can be described by the same three
numbers.  

A widespread approach to this problem is to let the sum of two such numbers be
equal to the sum of the central values, with the errors given by the upper and
lower errors added in quadrature \emph{separately}. That is, to set
\begin{eqnarray}
\label{eq:order0}
\nonumber
\X + \Y & \equiv & {x_0}_{-\sigloxs}^{+\sighixs} + {y_0}_{-\sigloys}^{+\sighiys}\\
        &      = & (x_0 + y_0)_{+\sqrt{\sigloxs^2 + \sigloys^2}}
                              ^{-\sqrt{\sighixs^2 + \sighiys^2}}.
                              \qquad\mathrm{(wrong!)}
\end{eqnarray}
This procedure has no statistical justification.  That it must be wrong can be
seen from considering the central limit theorem: In the limit of many
distributions of the same asymmetry, the combined PDF should approach a
Gaussian. In contrast, errors added according to \eq{order0} never decrease in
asymmetry.

Part of the confusion arises from the ambiguity in what represents ``the
central value''. A common interpretation is the most probable value of the
distribution, i.e.~the mode. %, indeed an intuitive thought e.g.~in the case of a
%PDF given by a thin peak and a long tail to one side.
An unwary scientist may then think that
$\mathrm{mode}[\X + \Y] = \mathrm{mode}[\X] + \mathrm{mode}[\Y]$.
But if \X\ and \Y\ are both skewed in the same direction, say towards negative
values such that $\siglo > \sighi$, the mode will be skewed in the same
direction.
%For instance, if $\siglo < \sighi$
If $x_0$ represents the mean $\mu$, the combined mean \emph{will} be the sum of
the two means, but errors added in quadrature will be even more off.  We argue
that, for asymmetric errors, a more natural estimator of the central value is
the median, and a natural estimator of a CI is the 68\% range that is given by
the 16th and 84th percentiles. In this case there is equal probability that a
number drawn from the PDF lies above and below the central value and, more
importantly, one can be sure that the central value lies between the lower and
the upper values, which is not necessarily the case for the mode and the mean
when dealing with asymmetric PDFs.

A further argument for using the $\{50,16,84\}$ percentiles is that --- in
contrast to other estimators --- they do not care whether the variable
in question in linearly or logarithmically distributed (for instance, the
median of an ensemble of measured fluxes picks out the same data point as the
median of the magnitudes).

A statistical error on $x$ can be thought of as deriving from a nuisance
parameter $a$. If $a$ itself has an uncertainty $\sigma_a$, then in the
``normal'', linear case that uncertainty propagates to $x$ as $\sigma_x^2 =
(dx/da)^2 \, \sigma_a^2$. Asymmetric errors arise when the relationship between
$\sigma_x$ and $\sigma_a$ is non-linear.  As mentioned above, if the full PDFs
of \X\ and \Y\ are not known, there is no ``correct'' solution to the problem.
Nevertheless, as will be described in the following it is possible to construct
a method that is a significant improvement over the (standard) quadrature
adding. This method is simply a numerical implementation of the technique
described by \citet{Barlow2003}. A Python function capable of performing such
additions can be downloaded from
\href{https://github.com/githyankipela/add_asym}
     {github.com/githyankipela/add\_asym}.

\subsection{Piecewise linear error propagation}
\label{sec:piece}

In the simplest non-linear approach, propagating the uncertainty from $a$ to
$x$ knowing only the set $\{x_0,\siglo,\sighi\}$ can be achieved assuming a
piecewise linear dependency. A similar method using a quadratic relationship
(not to be confused with ``addition of errors in quadrature'') has also been
implemented in the public code. There is no a priori reason to favor on over
the other; the piecewise linear transformation has an unphysical kink, while
the quadratic one has a turnover somewhere outside of the 68\% CI, but their
results are comparable nonetheless (and both much better than the ``standard''
method).

\begin{figure} %[!t]
\centering
\includegraphics [width=0.40\textwidth] {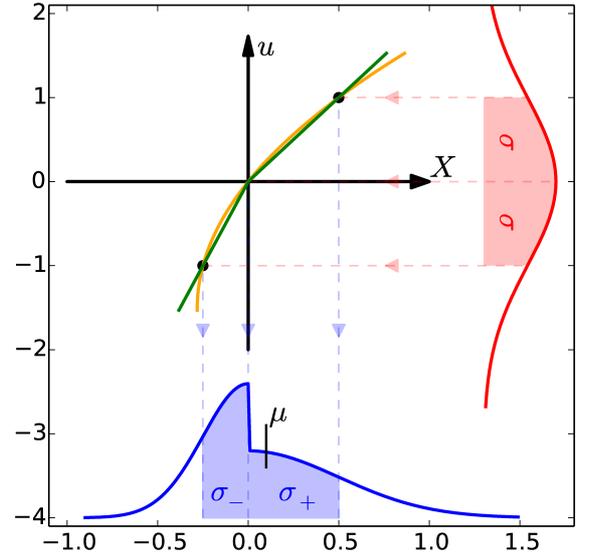}
\caption{Error propagation of a nuisance parameter $u$ (\emph{red} with
         transparent 1$\sigma$ confidence interval) to a variable $\xx$
         (\emph{blue} with 1\protect\siglo-1\protect\sighi confidence
         intervals) through a piecewise linear relationship (\emph{green
         line}). Here, $\protect\siglo=0.25$ and $\protect\sighi=0.50$. A
         quadratic relationship between $u$ and $X$ is also shown
         (\emph{orange}). The mean $\mu$ differs from the central value by a
         factor $(\protect\sighi-\protect\siglo)/\sqrt{2\pi} \simeq 0.1$.} 
\label{fig:addAsym}
\end{figure}

First, the nuisance parameter $a$ is transformed to a variable $u$ given by a
unit Gaussian, while as the dependent variable we will consider $\xx(u) = x(u)
- x(0)$. The function propagating the PDF of the nuisance parameter is
then
\begin{equation}
\label{eq:piece}
\xx = \left\{ \begin{array}{ll}
\siglo u & \textrm{for } u \le 0\\
\sighi u & \textrm{for } u \ge 0.
\end{array}
\right.
\end{equation}
A consequence of this transformation is that the expectation value, or the
average, of \xx\ no longer is equal to the most likely value, but instead given
by:
\begin{eqnarray}
\label{eq:mu}
\nonumber
\ave{\xx} = \mu & = & \int_{-\infty}^0 \! \! du\, \siglo u \frac{e^{-u^2/2}}{\sqrt{2\pi}}
                    + \int^{+\infty}_0 \! \! du\, \sighi u \frac{e^{-u^2/2}}{\sqrt{2\pi}}\\
                & = & \frac{1}{\sqrt{2\pi}} (\sighi - \siglo).
\end{eqnarray}
Thus, if $\xx(u)$ is a non-linear function of $u$, its expectation \ave{\xx} is
not $\xx(\ave{u})$, and if $\xx(0)$ is quoted as the central value, it is not
the mean. It will still be the median, though, and could as such still be
quoted as the central value.
The error propagation is illustrated in \fig{addAsym}.

With this transformation we obtain PDFs corresponding to the three-number set,
which can then be convolved with the PDF of another three-number set. But
convolving two non-Gaussians does not necessarily result in a PDF of the same
form as the original functions, as is the case for Gaussians. However, even
though the form is not preserved, some things are; the Thiele semi-invariant
cumulants \citep{Thiele1889} still add under convolution. The first three of
these are the usual mean (\eq{mu}), the variance
\begin{equation}
\label{eq:var}
V = \frac{1}{2}(\siglo^2 + \sighi^2) - \frac{1}{\sqrt{2\pi}} (\sighi - \siglo),
\end{equation}
and the unnormalized skewness
%%%%%%%%%%%%%%%%%%%%%%%%%%%%%%%%%%%%%%%%%%%%%%%%%%%%%%%%%%%%%%%%%%%
%   Decreases dist around =, but does so for the rest of the doc: %
%   {\setlength\arraycolsep{2pt}                                  %
%   Easier just to break manually                                 %
%%%%%%%%%%%%%%%%%%%%%%%%%%%%%%%%%%%%%%%%%%%%%%%%%%%%%%%%%%%%%%%%%%%
\begin{eqnarray}
\label{eq:gamma}
\gamma & = & \ave{x^3} - 3\ave{x}\ave{x^2} + 2\ave{x}^3\\
\nonumber
       & = & \frac{1}{\sqrt{2\pi}}
             \left[ 2(\sighi^3 - \siglo^3)
             - \frac{3}{2}   (\sighi - \siglo) (\sighi^2 + \siglo^2)
             \right. \\
\nonumber
       &   & + \frac{1}{\pi} (\sighi - \siglo)^3
             \left. \vphantom{\frac{3}{2}} \right].
\end{eqnarray}
Now the problem of adding $n$ PDFs has been reformulated so as to find the
distribution that has mean, variance, and unnormalized skewness equal to the
sums of the indvidual distributions:
$\{\mu_\mathrm{tot},V_\mathrm{tot},\gamma_\mathrm{tot}\} =
\{\sum_i^n\mu_i,\sum_i^n V_i,\sum_i^n\gamma_i\}$,
i.e. to revert Eqs.~\ref{eq:mu}, \ref{eq:var}, and \ref{eq:gamma}, which
express these quantities in terms of \siglo\ and \sighi. This must be
done numerically. Defining $D \equiv \sighi - \siglo$ and $S \equiv \siglo^2 +
\sighi^2$ the equations
\begin{eqnarray}
\label{eq:SD}
S & = & 2V + \frac{D^2}{\pi}\\
D & = & \frac{2}{3S} \left( \sqrt{2\pi}\gamma - D^3 (\frac{1}{\pi}-1) \right)
\end{eqnarray}
are solved iteratively, starting with $D = 0$ (takes only a handful of
iterations). The resulting PDF is then given by
\begin{eqnarray}
\label{eq:x0s1s2}
x_0    & = & \mu - \frac{D}{\sqrt{2\pi}}\\
\siglo & = & \bar{\sigma} - \frac{D}{2}\\
\sighi & = & \bar{\sigma} + \frac{D}{2}
\end{eqnarray}
where the (biased) mean is calculated through
\begin{equation}
\label{eq:munum}
\mu = \sum_i^n x_{0,i} + \frac{\sighi{_{,i}} - \siglo{_{,i}}}{\sqrt{2\pi}},
\end{equation}
and
\begin{equation}
\label{eq:sig}
\bar{\sigma} = \sqrt{V - \frac{D}{2} (1 - \frac{2}{\pi})}
\end{equation}
is the mean uncertainty.

%end piece

\subsection{Testing the scheme}
\label{sec:testAsym}

\begin{figure}%[!t]
\centering
\hspace*{-5mm}\includegraphics [width=0.55\textwidth] {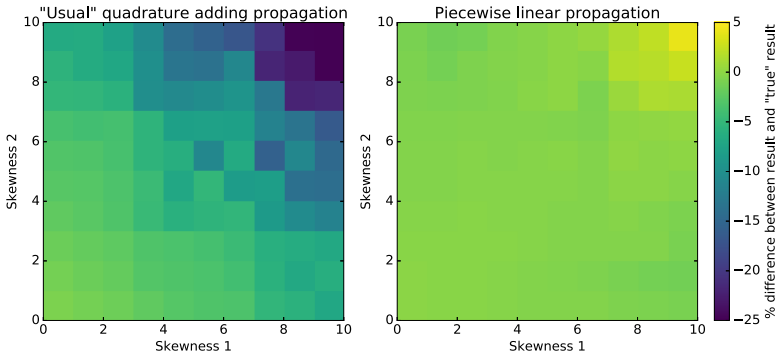}
\caption{Offset of the central value $x_0$ from the ``true'' value
         $x_{0,\mathrm{true}}$ obtained through convolution, in the case of the
         ``usual'', but wrong, method of adding errors in quadrature
         (\emph{left}), and a piecewise linear error propagation
         (\emph{right}), as a function of the skewness of the two addends.
         Here, offset is defined as $(x_0 - x_{0,\mathrm{true}}) /
         x_{0,\mathrm{true}}$. The former method is off by a large fraction for
         relatively small skewnesses, whereas the latter method is correct on
         the $<$1\% level even out to skewnesses of 8--9.}
\label{fig:testAsym}
\end{figure}

In order to test the method, we carry out a series of additions of PDFs.  The
tested functions are of various functional forms (lognormal, loglogistic,
Fr\'echet, and Weibull), and for each pair the value of their parameters are
drawn randomly in intervals such that their skewness ranges from 0 (symmetric)
to 10 (highly asymmetric). Their central values and lower and upper
uncertainties are calculated, and compared to their ``true'' values obtained
through a convolution of their full PDFs. The result is shown in \fig{testAsym}
for the central values of the ``usual'' (\emph{left}) and the piecewise linear
error propagation (\emph{right}) method.
The colour shows the offset of $x_0$ of the two methods from the true
result, defined as $(x_0 - x_{0,\mathrm{true}}) / x_{0,\mathrm{true}}$, as a
function of the normalized skewnesses of the two PDFs (defined as the third
moment about the mean, divided by the cube of the second moment about the
mean). The ``usual'' method causes an unwanted bias of $x_0$ of $>$1\% already
for small skewnesses of $\sim0.5$, and for moderate skewnesses of $\simeq5$ and
above, the bias is $>$10\%! This is for adding only two PDFs.; if multiple PDFs
are added, the bias accumulates, leading to unacceptably large inaccuracies. In
contrast, propagating errors through a piecewise linear transformation is
accurate on the $<$1\% level even out to large skewnesses of $\sim8$.

Similar figures can be constructed for the offset of \siglo\ and \sighi\, and
for the quadratic relationship. For the PDFs tested, the piecewise linear
method performs slightly better than the quadratic with regards to $x_0$, while
the opposite is true for \siglo\ and \sighi\ (with both methods performing much
better than the ``usual'' method). In this work,
%because errors are rather anyway,    <--- whaaaat?
%we are more concerned with the central value; hence
we have chosen to
apply the piecewise linear error propagation throughout.

%end testAsym
%end addAsym
\end{appendix}

% \bsp
% \label{lastpage}
\end{document}